\documentclass[a4paper,aps,pre,twocolumn,superscriptaddress,groupedaddress,floatfix]{revtex4}
\usepackage{graphicx}
\usepackage{amsmath}
\usepackage{natbib}
\usepackage{amssymb}
\usepackage{braket}
\usepackage{hyperref}
\usepackage{subfigure}
\usepackage{array}
\usepackage{float}
\usepackage[usenames,dvipsnames]{color}
\usepackage{graphicx}
\usepackage{subfigure}
\usepackage{float}
\usepackage{verbatim}
\usepackage{cleveref}
\usepackage{pdfsync}
\usepackage[version=4]{mhchem}
\usepackage{chemarr}
\usepackage[export]{adjustbox}
\newcommand{\qd}{\, .}
\newcommand{\qc}{\, ,}
\newcommand{\gd}{\phi_{i}}

\newcommand{\bc}{k_{B}}

\newcommand{\gpf}{P_{i}}

\newcommand{\tot}{N}

\newcommand{\dt}{\partial_{t}} 
\newcommand{\flux}{\vect{{\textit{\textbf{j}}}}} 
\newcommand{\var}{\textit{\textbf{r}},t}
\newcommand{\varx}{\textit{\textbf{r}}}

\newcommand{\avgvf}{\bar{\phi}}

\newcommand{\vect}[1]{\boldsymbol{#1}}

\newcommand{\be}{\begin{equation}}
\newcommand{\ee}{\end{equation}}
\newcommand{\III}{\text{I/II}}
\newcommand{\I}{\text{I}}
\newcommand{\II}{\text{II}}

\newcommand{\cpref}{\bar{\mu}^0}
\begin{document}
\begin{@twocolumnfalse}
\title{
Chemical Reactions regulated by Phase-Separated Condensates}

\author{Sudarshana Laha$^{1,2}$, Jonathan Bauermann$^{1,2}$, 
Frank J\"ulicher$^{1,2,4}$, Thomas C.T. Michaels$^{5,6}$ and Christoph A. Weber$^{1,2,7,8}$}
\affiliation{$^{1}$Max Planck Institute for the Physics of Complex Systems,
Nöthnitzer Strasse~38, 01187 Dresden, Germany\\$^{2}$Center for Systems Biology Dresden,  Pfotenhauerstrasse~108, 01307 Dresden, Germany\\
$^{4}$Cluster of Excellence Physics of Life, TU Dresden, 01062 Dresden, Germany\\
$^{5}$Department of Biology, Institute of Biochemistry, ETH Zurich, Otto Stern Weg 3, 8093 Zurich, Switzerland\\
$^{6}$Bringing Materials to Life Initiative, ETH Zurich, Switzerland\\
$^{7}$Faculty of Mathematics, Natural Sciences, and Materials Engineering: Institute of Physics, University of Augsburg, Universit\"atsstr. 1, 86159 Augsburg, Germany\\
$^{8}$Corresponding author: {christoph.weber@physik.uni-augsburg.de}}

\begin{abstract}
Phase-separated liquid condensates can spatially organize and thereby regulate chemical processes. However, the physicochemical mechanisms underlying such regulation remain elusive as the intramolecular interactions responsible for phase separation give rise to a coupling between diffusion and chemical reactions at non-dilute conditions. Here, we derive a theoretical framework that 
decouples the phase separation of scaffold molecules from the reaction kinetics of diluted clients. As a result, phase volume and client partitioning coefficients become control parameters, which enables us to dissect the impact of phase-separated condensates on chemical reactions. 
We apply this framework to two chemical processes and show how condensates affect the yield of reversible chemical reactions and the initial rate of a simple assembly process. 
In both cases, we find an optimal condensate volume at which the respective chemical reaction property is maximal. Our work can be applied to experimentally quantify how condensed phases alter chemical processes in systems biology and unravel the mechanisms of how biomolecular condensates regulate biochemistry in living cells.
\end{abstract}
\maketitle	
\end{@twocolumnfalse}

\section{Introduction}\label{intro}

Living cells are spatially organized by compartments such as organelles~\cite{Wilson:1899} and protein-RNA condensates~\cite{Hyman:2014, Shin:2017}. 
While organelles like mitochondria are enclosed by membranes,  protein-RNA condensates lack a membrane and are instead phase-separated biomolecular condensates that coexist with the cyto- or nucleoplasm~\cite{Brangwynne:2009,Feric:2016,Boeynaems:2018,Antifeeva:2022}. 
Both types of compartments provide specific physicochemical environments that are required for the occurrence of various chemical reactions and related biological functions~\cite{Banani:2017,Alberti:2017}. 
While membrane-bound organelles use active membrane pumps to create such specific environments~\cite{PEDERSEN:1987,Beyenbach:2006,HEALD:2014}, phase-separated condensates differ in their composition to the outside already at phase equilibrium~\cite{Hyman:2014,DITLEV:2018}. 
Owing to this composition difference, reacting components partition differently between the phases~\cite{Frankel:2016,WOODRUFF:2017,AUMILLER:2017}. As a result, diffusion coefficients and reaction rate coefficients are distinct to each  phase~\cite{Elbaum-Garfinkle:2015,Lars:2021}. The phase coexistence between a liquid condensate and its surrounding phase was suggested to regulate various chemical reactions in living cells and model systems in the field of systems biology, 
such as protein aggregation and phosphorylation~\cite{Strulson:2012,Sokolova:2013,Alberti:2017,Drobot:2018,nakashima:2018,arosio:2021,O’Flynn:2021,Schoenmakers:2023}. 

However, when chemical reactions occur, they can give rise to diffusive fluxes through the condensate interface separating the phases of different composition~\cite{weber2019physics,Bo:2021, zwicker2022intertwined, BauermannLaha:2022, Bauermann:2022}. 
Such diffusive fluxes affect phase coexistence, i.e., their volumes and compositions. 
These changes, in turn, create feedback on the kinetics of chemical reactions. 
This mutual coupling between chemical and diffusive fluxes makes it difficult to dissect the effects of phase-separated condensates on chemical reactions and establish generic underlying principles.

In biology, a terminology was established for proteins that form biomolecular condensates in living cells~\cite{Banani:2017, ditlev2018s,gao2022brief}. 
Such proteins were assigned to two classes: scaffolds or clients. 
The scaffold components are thought to be the main components that ``create'' the condensates, while the clients can simply ``visit'' the condensate by partitioning without significantly affecting the properties of the condensate. 
Clients typically participate in chemical processes that are associated with biological functions~\cite{Banani:2017, woodruff2018organization, Lyon:2021}. 
Strictly speaking, this class assignment is, however, not possible in general due to the mutual coupling between phase separation and chemical processes.

\begin{figure}[tb]
\centering
\includegraphics[width=1.0\columnwidth]{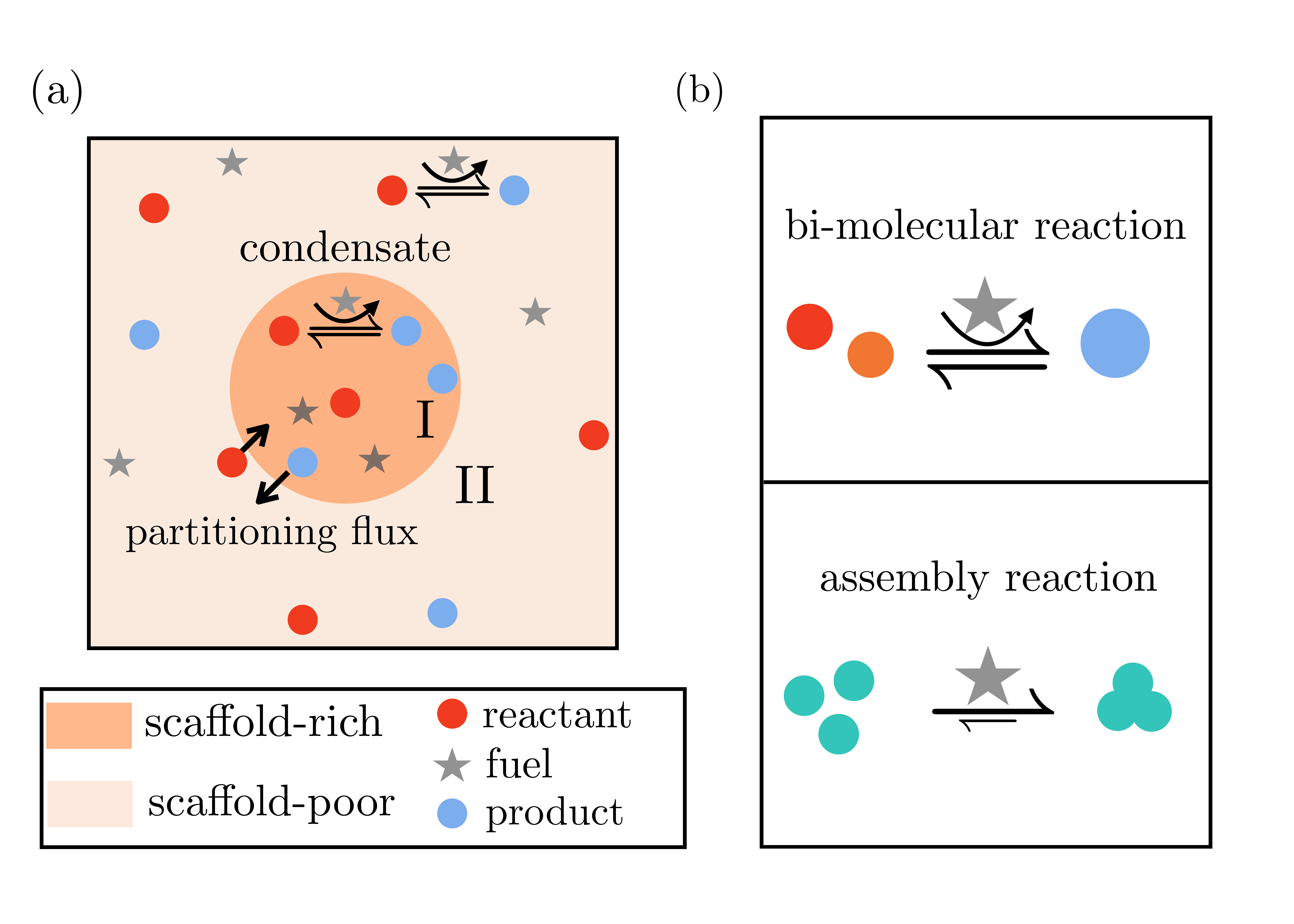}
\caption{\textbf{Schematic representation of a system with dilute reacting clients in the presence of a single spherical condensate.} 
\textbf{(a)} The non-dilute scaffold component can form a spherical condensate (phase I). The dilute clients (reactant and product) partition between the condensed, scaffold-rich phase and the scaffold-poor phase and undergo chemical reactions in each phase indicated as I and II. In addition, clients are exchanged between two phases via diffusion, maintaining the partition equilibrium of clients.
\textbf{(b)} Overview of the different considered chemical processes. We study the kinetics of each reaction with and without a condensate to decipher how phase coexistence affects chemical reactions.
We distinguish between reactions that can relax toward chemical equilibrium and reactions that are driven out of equilibrium by the continuous supply of fuel energy, indicated by star symbols in (a) and (b).}
\label{fig_1}
\end{figure}

To bridge this gap, we developed a theoretical framework to describe the chemical reactions of diluted clients within phase-separated condensates. 
The diluted clients interact with non-dilute scaffolds and solvents that phase-separate into a condensate rich in scaffolds that coexist with its surrounding scaffold-poor phase. 
However, consistent with the class assignment used in cell biology, the clients cannot affect the properties of the condensates.  
To illustrate the effects of condensates on chemical processes, we consider simple but biologically relevant chemical processes.
In particular, we calculate the yield of reversible chemical reactions and the initial rates of a simple assembly process 
(see Fig.~\ref{fig_1}(a,b)). 
Our key finding is that reaction yields and initial rates can be maximal for distinct condensate volumes. 
The possibility of tuning the system to this maximum suggests the relevance of condensate size for the control of specific chemical reactions in living cells. 

\section{Theoretical framework for reacting clients in coexisting phases}\label{model}

In this section, we discuss a general theoretical framework for chemical reactions regulated by liquid-like, phase-separated condensates in an incompressible system~\cite{Bauermann:2022, BauermannLaha:2022}. 
This framework is derived for diluted components that can undergo chemical reactions, which we refer to as clients in the following. 
These clients are diluted relative to a scaffold component and the solvent. 
The client volume fractions are non-zero such that chemical reactions among clients can occur but are small compared to the volume fractions of the scaffold and the solvent. 
The scaffold and solvent are non-dilute and considered not to participate in the chemical reactions among the clients; for simplicity, we neglect chemical reactions between scaffold and solvent. 
The condensate is rich in this scaffold component and labeled by I. 
The condensate phase coexists with the scaffold-poor phase II. 
For simplicity, we focus on one spherical condensate embedded in a spherically symmetric finite system; see Fig.~\ref{fig_2}(a). 
Using our framework, we will learn that clients' kinetics depend on the coexisting phases, while the latter remains unaffected by the clients.

\subsection{General theoretical framework}

We consider an incompressible mixture of $(\tot+2)$ components including  a non-reacting solvent ($i=0$), a non-reacting scaffold ($i=1$) and $\tot$ reacting clients.
Incompressibility corresponds to the case where each molecular volume $\nu_i$ is constant.
It implies that
we can describe the mixture's composition by $(\tot+2)$ volume fraction fields $\phi_i(\textbf{x},t)$ with $i=0,..., (\tot+1)$, where $\textbf{x}$ denotes position and $t$ time.
Such volume fraction fields obey  $\phi_0=1-\sum_{i=1}^{\tot+1} \phi_i$, which allows us to substitute the solvent volume fraction $\phi_0$ in the following framework.
The time evolution of the volume fraction fields $\phi_{i}(\textbf{x},t)$ of the remaining $(\tot+1)$ components  ($i=1,..., (\tot+1)$) reads:
\begin{subequations}\label{eq:kinetic_eq}
\be
\dt \phi_{i}=-\nabla \cdot \textbf{j}_{i}+ s_{i} \,,
\ee
with the diffusive fluxes using linear response~\cite{Frank_JProst2008} given as: 
\be \label{eq:fluxes_linres}
\textbf{j}_{i}=-\sum_{k}M_{ik}
\nabla \bar{\mu}_{k}  
\, .
\ee
\end{subequations}
Here, $\bar{\mu}_{i}$ is the exchange chemical potential of  component $i$, each calculated relative to the solvent.
Moreover, $M_{ik}$ is the volume fraction-dependent, non-diagonal, and, in general, non-symmetric mobility matrix. 
The non-diagonal terms in $M_{ik}$ are called cross-coupling coefficients.
Since the concentration of component $i$, $\nu_i \phi_i$ with $\nu_i$ denoting its molecular volume, is the conjugate thermodynamic quantities to the exchange chemical potential $\bar{\mu}_i$,  the symmetric Onsager matrix is $M_{ik}/\nu_i$ with Onsager's reciprocal relationship  
$M_{ik}\nu_k=M_{ki}\nu_i$. 

The exchange chemical potentials
can be expressed as follows~\cite{Bauermann:2022}:
\begin{align}\label{eq:chem_pot}
\bar{\mu}_{i}(\{\phi_{j}\})&=\bar{\mu}^{0}_{i} +k_{B}T\log(\bar{\gamma}_{i}(\{\phi_{j}\})\phi_{i})-\kappa_i  \nabla^2 \phi_i \, ,
\end{align} 
where $\bar{\mu}^{0}_{i}(T)$ is the exchange reference chemical potential, $\bar{\gamma}_i(\{\phi_{j}\}, T)$ denotes the exchange activity coefficient and $T$ is temperature.
The term proportional to $\log{\phi_i}$ originates from the mixing entropy (Appendix~\ref{sect:app_thermodynamics_NPLUS2COMPMIX}). 
While $\bar{\mu}^{0}_{i}(T)$ includes component-specific internal free energies and a shift with respect to normal conditions, 
the exchange activity coefficient 
$\bar{\gamma}_i(\{\phi_{j}\}, T)$ contains the interactions between component $i$ and $j$.
These exchange activity coefficients describe how phase separation of the scaffold component affects the chemical kinetics and the difference to the dilute mass action law in homogeneous systems. 
Moreover, $\kappa_i$ characterizes the free energy penalties for gradients in volume fractions $\phi_i$, where we have omitted cross-couplings for simplicity; see Appendix~\ref{sect:app_thermodynamics_NPLUS2COMPMIX} for a more detailed discussion.

The form of the mobility matrix $M_{ik}$ in Eq.~\eqref{eq:fluxes_linres} can be
obtained by the following argument: 
In the dilute limit of all components $i=1,...,(\tot+1)$ relative to solvent $i=0$, the diffusion matrix 
$D_{il}=\sum_k M_{ik} \partial \bar{\mu}_k/\partial \phi_l$ 
in $\vect{j}_i = - \sum_l D_{il} \nabla \phi_l + M_{ii} \kappa_i \nabla \nabla^2 \phi_i$ is independent of composition. 
Thus, the mobility matrix $M_{ik}$ has to cancel the composition dependence arising from the entropic  contribution of the exchange chemical potential gradient, $-T\partial( \log{\phi_i})/\partial \phi_i$ (Eq.~\eqref{eq:chem_pot}).
If we additionally aim for a diagonal diffusion matrix without any cross-diffusion
in the dilute limit of all components $i=1,...,N+1$, i.e., $D_{ii}=m_{0i}k_BT$, we can choose the following form for the mobility matrix~\cite{Bo:2021}: 
%
\be
\begin{split}\label{eq:mob_matrix}
    & M_{ii}= m_{0i}\phi_{i}(1-\phi_{1}-\sum^{\tot+1}_{i=2}\phi_{i})+ \sum^{\tot+1}_{\substack{k\neq i \, , \\k=1}}m_{ik}\phi_{i}\phi_{k}\, , \\
    & M_{ik} = -m_{ik}\phi_{i}\phi_{k}\, , \,  \forall i\neq k \, ,
\end{split}
\ee
with $m_{ik}$ being mobility coefficients that obey due to Onsager reciprocal relationship, $m_{ik}\nu_k = m_{ki}\nu_i$. 

The general reaction scheme between the chemical components $C_i$ can be written as
\be 
\sum_{i=2}^{\tot+1} \sigma_{i\alpha}^+ C_i \rightleftharpoons \sum_{i=2}^{\tot+1} \sigma_{i\alpha}^- C_i \qc \label{eq:chem_scheme}
\ee
where the $ \sigma_{i\alpha}^{\pm}$ are the stoichiometric coefficients for the reactants and the products, respectively. 
Moreover, $\alpha=1, ..., R$ labels the chemical reaction, and $R$ is the total number of chemical reactions in the system. 
The chemical reaction rates for the reactive clients, $s_{i}$, is given as follows~\cite{Bauermann:2022}:
\begin{subequations}\label{eq:prod_rate_with_H}
\be
s_i = \nu_i\sum^{R}_{\alpha=1} k_\alpha \sigma_{i\alpha}  H^{}_{\alpha} 
\qc
\ee
where 
$\sigma_{i\alpha}=\sigma_{i\alpha}^{-}-\sigma_{i\alpha}^{+}$, 
 $k^{}_{\alpha}$ is the composition-dependent reaction rate coefficient of reaction $\alpha$ which can differ between the phases $\III$, $\nu_i$ is the molecular volume of component $i$ and $V$ is the system volume. 
As the system is incompressible, the chemical reactions have to conserve volume, i.e., 
$\sum_{i}  s_i=0$ implying that 
$\sum_{i} \sigma_{i\alpha} \nu_i = 0$ for each chemical reaction $\alpha$. 
Moreover, $H^{}_{\alpha}$ is the reaction force which can be expressed as follows:
\be
\label{eq:reaction_force_b}
\begin{split}
H^{}_{\alpha} &=  \prod^{\tot+1}_{i=2} \left[\exp{\left(\frac{\bar{\mu}_i^0+\tilde{\mu}^{+}_{F}-\kappa_i \nabla^2 \phi_i}{\bc T}\right)} \bar{\gamma}_{i} \phi_{i}^{}\right]^{\sigma_{i\alpha}^+} \\
&\quad- 
\prod^{\tot+1}_{i=2} \left[ \exp{\left(\frac{\bar{\mu}_i^0+\tilde{\mu}^{-}_{F} -\kappa_i \nabla^2 \phi_i }{\bc T}\right)}\bar{\gamma}_{i}^{}\phi_{i}^{}\right]^{\sigma_{i\alpha}^-} \qd
\end{split}
\ee
\end{subequations}
Here, the first terms correspond to the gain contributions for the products, while the second terms are the product's loss terms. 
Moreover, $\tilde{\mu}^{\pm}_{F}$ denotes a fuel energy supply that maintains the system away from equilibrium. 
This can be achieved by differing $\tilde{\mu}^{\pm}_{F}$, between the phases $\III$.
Such a case  could be realized by chemostats for fuel components and their waste to which the fuel turns over; see Ref.~\cite{Bauermann:2022} for a more detailed discussion.  
Without fuel energy supply ($\tilde{\mu}^{\pm}_{F}=0$) and when the system is at phase equilibrium ($\bar{\mu}_{i}^\I=\bar{\mu}_{i}^\II$), 
the reaction force 
$H^{\III}_{\alpha}$ is phase-independent~\cite{BauermannLaha:2022}.

\subsection{Theoretical framework  for diluted clients}

\subsubsection{Continuum description}

When all clients $i=2,..., (\tot+1)$ are diluted relative to the solvent ($i=0$) and scaffold ($i=1$) components, their exchange activity coefficients, defined in Eq.~\eqref{eq:chem_pot}, approach constant values for small client volume fractions. Such values solely depend on the scaffold equilibrium volume fraction $\phi_{1}$ but not on the volume fraction of other clients $\{\phi_j\}$.
For a mean-field free energy including a mixing entropy and interactions up to second order in volume fractions (Eq.~\eqref{eq:Ap_freeenergy}), the exchange activity coefficients defined via Eq.~\eqref{eq:chem_pot} read for diluted clients 
(derivation see Appendix~\ref{App:1}):
\be\label{eq:bargammaclient}
\bar{\gamma}_{i}=\frac{1}{(1-\phi_{1})^{r_i}}\exp{\bigg[r_{i}\phi_{1}(\chi_{1i}-\chi_{0i}-\chi_{01})\bigg]} \, ,
\ee
where $r_i=\nu_{i}/\nu_{0}$ denotes the ratio of the $i$-th component's molecule volume $\nu_{i}$ relative to the solvent molecular volume $\nu_{0}$.
Moreover, the strength of molecular interactions among the components is characterized by interaction parameters, i.e., between scaffold and solvent $\chi_{01}$, scaffold with clients $\chi_{1i}$, and solvent with clients $\chi_{0i}$. The client-client interactions do not appear in Eq.~\eqref{eq:bargammaclient} since clients are dilute with respect to solvent and scaffold. 
When in addition, the volume fraction of the scaffold becomes small ($\phi_1 \to 0$ in Eq.~\eqref{eq:bargammaclient}), all exchange activity coefficients of clients approach unity. 
Note that Eq.~\eqref{eq:bargammaclient} 
is also valid when considering free energies beyond mean-field with interaction terms beyond second order in volume fractions. The reason is that it already captures  the leading order 
coupling between clients with scaffold and solvent components (see Appendix~\ref{App:1}).

For diluted clients in a system with non-dilute scaffold and solvent, $\sum^{\tot+1}_{i=2}\phi_{i} \ll \phi_1$, cross-couplings between 
clients and scaffolds/solvent vanish ($M_{1i}=0, i>1$), and the cross-couplings between clients are negligible ($M_{ij}=0, i,j>1$). 
Thus, 
the mobility matrix (Eq.~\eqref{eq:mob_matrix}) becomes for diluted clients ($i>1$): 
\begin{subequations}\label{eq:mob_matrix_clients}
\begin{align}
& M_{11} = m_{01}\phi_{1}(1-\phi_{1})
\, , \\
& M_{1i} = 0  \, ,  \\
& M_{ii} = m_{0i}\phi_{i}(1-\phi_{1})+m_{1i}\phi_{i}\phi_{1} \, ,  \\
& M_{i1} = 0\, ,    \\
& M_{ij} = 0 \, ,  \quad i\not= j>1\, . \quad 
\end{align}
\end{subequations}
Using Eq.~\eqref{eq:kinetic_eq}, the time evolution of the scaffold volume fraction follows a Cahn-Hilliard equation for a binary mixture~\cite{Bray:1993}:
\begin{align}
\nonumber
\dt \phi_{1} &= \nabla \cdot \bigg[ m_{01}\phi_{1}(1-\phi_{1})\bigg(k_{B}T \left(\frac{1}{\phi_{1}}+\frac{1}{\bar{\gamma}_{1}}\frac{\partial \bar{\gamma}_1}{\partial \phi_{1}}\right)\nabla \phi_{1}\\
\label{eq:sc_cm}
&\quad -\kappa_1\nabla\nabla^{2}\phi_{1} \bigg) \bigg] \, .
\end{align}
The kinetic equations for the clients $i=2,..., (\tot+1)$ are given by:
\begin{subequations}\label{eq:c_cm}
\be
\dt \phi_{i} =-\nabla \cdot\textbf{j}_{i} +s_{i} (\{\phi_{j}\}) \, , 
\ee
with the reaction rate $s_{i} (\{\phi_{j}\})$ and the client flux
\be
\label{eq:client_flux}
\textbf{j}_{i} =-\textbf{\emph{v}}_i(\phi_{1},\nabla\phi_{1})\phi_{i}-D_i(\phi_{1})\nabla\phi_{i}
\, . 
\ee
We have neglected the contribution to the client flux
$\kappa_i \nabla \nabla^2 \phi_i$  as the spatial transport of clients is well-captured by the leading order diffusive flux that is proportional to $\nabla \phi_i$; see Appendix~\ref{ap:cont} for more details. 

Due to phase separation between scaffold components and solvent, clients are effectively subject to a drift velocity $\textbf{\emph{v}}_i(\phi_{1},\nabla\phi_{1})$.
This effective drift arises from cross-diffusion where clients are driven
by gradients in scaffold components, $\nabla\phi_{1}$~\cite{Bo:2021}. Moreover, clients diffuse with diffusion coefficients $D_i(\phi_1)$ that is set by the local scaffold volume fraction  (derivation see Appendix~\ref{ap:cont}):
\begin{align}
\label{eq:drift_coeff_cont}
\textbf{\emph{v}}_i(\phi_{1},\nabla\phi_{1})&=
 D_i(\phi_{1}) \frac{1}{\bar{\gamma}_i} \frac{\partial \bar{\gamma}_i}{\partial \phi_1}
\nabla\phi_{1} \, ,
\\
\label{eq:diffusion_ceoff_cont}
  D_i(\phi_{1}) &= k_{B}T\bigg[m_{0i}(1-\phi_{1})+m_{1i}\phi_{1}\bigg] \, .
\end{align}
\end{subequations}
The effective drift velocities $\textbf{\emph{v}}_i$ depend on the local diffusion coefficients $D_i(\phi_1)$ highlighting once more their origin in cross-diffusion.  
The effective drift velocities also depend on the thermodynamic parameters such as molecular volumes $\nu_i$, encoded in $r_i=\nu_i/\nu_0$, and the interaction parameters $\chi_{1i}$ and $\chi_{0i}$. These dependencies come from the activity coefficients in Eq.~\eqref{eq:drift_coeff_cont} via: 
\begin{equation}
  \frac{1}{\bar{\gamma}_i} \frac{\partial \bar{\gamma}_i}{\partial \phi_1} = 
  r_{i}(\chi_{1i}-\chi_{0i}-\chi_{01}) + \frac{r_i}{\left({1-\phi_1^{}}\right)} \, .
\end{equation}
Note that the limit $\phi_1 \to 1$ does not lead to a divergence of the drift velocity because  the system becomes homogeneous and  $\nabla \phi_1$ vanishes. 

Our derivation shows that the time-dependent spatial profiles of the  scaffold $\phi_1(\textbf{x},t)$ determine a spatiotemporal environment for clients undergoing diffusion and chemical reactions. 
Note that the chemical reaction rates $s_i$ for diluted clients are given by Eq.~\eqref{eq:prod_rate_with_H} with the exchange activity coefficients $\bar{\gamma}_{i}$ shown in Eq.~\eqref{eq:reaction_force_b} that solely depend on the scaffold volume fraction. The term $\kappa_i \nabla^2 \phi_i$ in the reaction force $H_\alpha$ can be neglected as it solely creates a positive up-shift of the diffusion coefficient (to lowest order when expanding the exponential in Eq.~\eqref{eq:reaction_force_b}). 

\begin{figure}[tb]
\centering
\includegraphics[width=1.0\columnwidth]{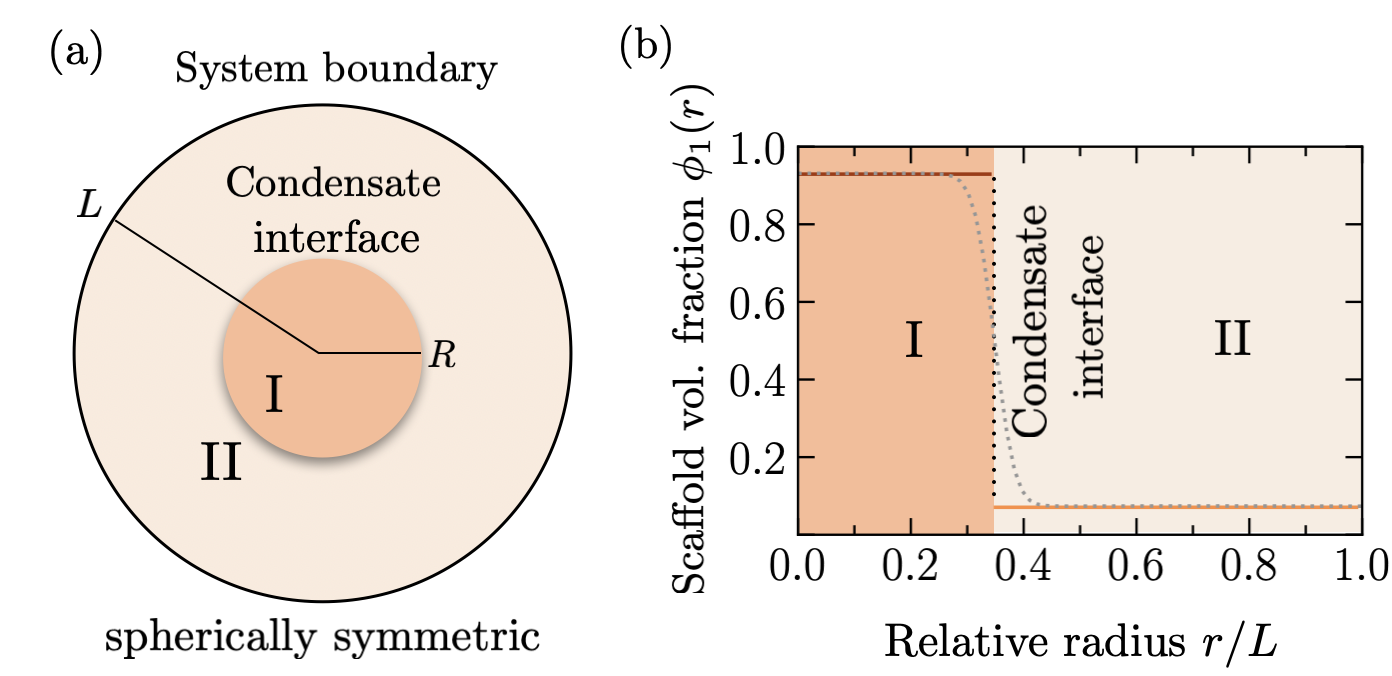}
\caption{\textbf{Schematic representation of the system to study how phase coexistence affects chemical reactions of dilute clients.} 
\textbf{(a)} We consider a spherical, condensate of radius $R$, which is rich in scaffold components (phase $\I$) coexisting with a scaffold-poor phase (phase $\II$). The system is also spherical with radius $L$.
\textbf{(b)} Stationary spatial profiles of the scaffold $\phi_{1}(r)$ in the dilute limit of clients; continuum model (grey dotted) and thin interface model (solid orange).}
\label{fig_2}
\end{figure}

A key hallmark of the chemical kinetics in a mixture that is composed of the scaffold, solvent, and diluted clients is that the scaffold volume fraction evolves according to Eq.~\eqref{eq:sc_cm} completely independent of the client volume fraction. In other words, the clients do not affect the phase separation between the scaffold component and solvent. 
However, the dynamic equations~\eqref{eq:c_cm} for the clients depend on the scaffold volume fraction. 
Once the scaffold volume fraction has settled in a stationary state ($\partial_{t} \phi_1=0$ in Eq.~\eqref{eq:sc_cm}), it exhibits a sigmoidal profile with the plateaus equal to the scaffold volume fractions at phase equilibrium $\phi^\text{\III}_{1}$; see Fig.~\ref{fig_2}(b). 
The respective plateau values determine the kinetic coefficients of clients, such as reaction rate coefficients and diffusivities, in phase I and II.
For a radially symmetric system with the radial position $r$, this stationary profile 
is approximately given as~\cite{weber2019physics}
\be \label{eq:tanh}
\phi_{1}^\text{int}(r)\simeq \frac{1}{2}(\phi^\text{I}_{1}+\phi^\text{II}_{1}) -\frac{1}{2}(\phi^\text{I}_{1}-\phi^\text{II}_{1})\tanh{\left((r-R)/\lambda^{\text{int}}\right)}\qc
\ee
where the interface width is 
\begin{equation}
\lambda^{\text{int}}=\sqrt{\frac{\kappa}{(\chi_{01}-2)k_{B}T \nu_{0} r_{1}}}\, .\end{equation}
For simplicity, we considered the limit $\lambda^{\text{int}} \ll R$, allowing us to neglect the effects related to the Laplace pressure and the curvature of the condensate~\cite{Bray:1993}.

\subsubsection{
Stationary scaffold
in thin interface model}\label{sect:thin_interface}

Here, we discuss an approximation of the continuous model for a  scaffold volume fraction profile that is stationary ($\partial_t\phi_1=0$ in Eq.~\eqref{eq:sc_cm}) with an interface width that is thin compared to the scaffold-rich and poor bulk phases, respectively, i.e.,  $\lambda^\text{int}/R \rightarrow 0$. 
In this limit, the continuous profile given by Eq.~\eqref{eq:tanh} becomes a step profile that jumps at the interface  (Fig.~\ref{fig_2}(b); \cite{elder2001sharp}). 
Stationarity of the scaffold profile that is independent of the dilute clients implies that $\phi_1$ is constant in each phase. 
Thus, the effective drift velocity  
$\textbf{\emph{v}}_i(\phi_{1},\nabla\phi_{1})$ and 
higher order contributions to the client flux,
$\nabla \nabla^{2}\phi_{1}$,  vanish  in  Eq.~\eqref{eq:client_flux}.
In the following, we consider 
a single, spherical condensate of radius $R$ in a radially symmetrical, spherical domain of diameter $2L$ ($\varx=(r,\theta,\varphi)$ denote the spherical coordinates).

Away from the interface and for a stationary scaffold profile $\phi_1(\vect{r})$, 
the dynamical equation for the dilute reacting clients in both phases $\III$ correspond to reaction-diffusion equations:
\begin{subequations}\label{eq:reacdiffbasic}
\be\label{eq:reacdiff1}
\dt \gd^{} (\var)= D^{\III}_{i} \nabla^{2} \gd^{} (\var) + s^{\III}_{i} (\{\phi_{j}(\var)\}) \, ,
\ee
where index $i=2,..., (\tot+1)$, represents the individual clients in the system. 
Here, phase I is located in $0<r<R$, while phase II extends in the range, $R<r<L$.
Moreover, $s_{i}^\III$ are the chemical reaction rates for the $i$-th client, which is given in Eq.~\eqref{eq:prod_rate_with_H} with reaction rate coefficients $k_\alpha^\III$ and fuel energy supply $\tilde{\mu}^{\pm,\III}_{F}$ being phase-dependent.
Note that Eq.~\eqref{eq:reacdiff1} is obtained 
without expanding around the phase equilibrium volume fractions~\cite{weber2019physics, Bauermann:2022}. 
Client diffusion following Fick's law combined with, in general, non-linear chemical reaction rates is a consequence of clients being diluted with respect to the phase-separated scaffold and solvent components. 
Furthermore, since the scaffold volume fraction $\phi_1^\III$ is homogeneous in each phase, the diffusion coefficients of clients are constants that solely differ among the phases:
\begin{align}
D^{\III}_{i} = 
\begin{cases}
k_{B}T\bigg(m_{0i}(1-\phi^{\I}_{1})+m_{1i}\phi^{\I}_{1}\bigg)\, & \text{for} \, \, 0<{r} <R\,  \, , \\
k_{B}T\bigg(m_{0i}(1-\phi^{\II}_{1})+m_{1i}\phi^{\II}_{1}\bigg) \, & \text{for} \, R<{r}<L \, . 
\end{cases}
\end{align}
Note that the client diffusion constants $D^{\alpha}_{i}$ are independent of client concentrations. Thus, they remain constant during the chemical reaction kinetics of the clients, while client profiles vary in time and space.

The reaction-diffusion equations of each component $i$ (Eq.~\eqref{eq:reacdiff1}) are coupled via the boundary conditions at the interface. 
At the interface $r=R$ and the system boundary $r=L$ (Fig.~\ref{fig_2}(a)), we can write the following boundary conditions for the clients' dynamic Eq.~\eqref{eq:reacdiff1}:
\begin{itemize}
\item Volume conservation of clients across the interface implies that the radial fluxes across the interface, $\vect{e}_r \cdot \flux_{i}^{}$, inside ($r=R_{-}$) and outside ($r=R_{+}$) of the interface are equal, 
\be
\label{eq:bc_conservation_interface}
 \vect{e}_r \cdot \flux_{i}^{}|_{r=R_{-}} = \vect{e}_r \cdot  \flux_{i}^{}|_{r=R_{+}} \, ,
 \ee
 where
 $\flux_{i}^{}=- \vect{e}_r \, D_i^\alpha \partial_r \phi
_i^{} $ is the flux and $\vect{e}_r$ denotes the radial unit vector.
\item Local phase equilibrium at the interface leads to client volume fractions inside and outside of the interface that satisfy the partition coefficient, 
\be\label{eq:bc_partitioning_interface}
	\gpf=\frac{\gd^{\text{}}|_{r=R_{-}}}{\gd^{\text{}}|_{r=R_{+}}}\, .
\ee
For diluted clients, the partition coefficients are constant and are determined by the interaction strength of clients with scaffold and solvent components, as well as the volume fraction difference of scaffold between inside and outside.
\item For a system boundary at $r=L$ that is impermeable for the clients, 
the diffusive flux of clients vanishes: 
\be\label{eq:bc_noflux_L}
	\vect{e}_r \cdot \flux_{i}^{} |_{r=L}=0 \, .
\ee

\item In the center of the spherical condensate at $r=0$, the flux has to vanish for each client: 
\be\label{eq:bc_noflux_center}
	\vect{e}_r \cdot \flux_{i}^{} |_{r=0}=0 \, .
\ee	
\end{itemize}
\end{subequations}

A solution of the volume fraction profile of a client at different time points is shown in Fig.~\ref{fig_3uni}(a), where the jump at the interface is set by its partition coefficient. The details of the considered chemical processes are discussed in Sect.~\ref{sec:App_chem}.

\subsubsection{Model at  phase equilibrium}\label{sect:phase_eq_section}

In this section, we discuss the case when clients diffuse fast compared to their reaction, i.e.,  the reaction rate coefficients corresponding to the linearized reaction rates are small compared to the system's slowest diffusion rates, $D_{i}^\III/L^2$.
In this case, the 
phases are homogeneous and at phase equilibrium with respect to each other at all times during the chemical kinetics. 
This limiting case reduces the mathematical complexity of the theoretical description significantly, and chemical kinetics is governed by ordinary differential equations~\cite{weber2019spatial, michaels2022enhanced, BauermannLaha:2022}, similar to the classical mass action law kinetics in homogeneous systems.

At phase equilibrium, the client volume fractions  $\phi^\III_i$ in each phase $\III$  are homogeneous and satisfy the partition coefficient, 
\begin{equation}\label{eq:partitioning_coeff}
    P_i=\frac{\phi^\I_i}{\phi^\II_i}  \, . 
\end{equation}
Using the exchange chemical potential (Eq.~\eqref{eq:chem_pot}), the partitioning coefficients can be expressed at phase equilibrium in terms of exchange activity coefficients, $P_i= \bar{\gamma}_{i}^\II/\bar{\gamma}_{i}^\I$. 
For the discussed case of diluted clients (Eq.~\eqref{eq:bargammaclient}), the client partitioning coefficients are constants and, thus, control parameters. 

At phase equilibrium, the volume fractions of the clients in the two phases $\I$ and $\II$, 
\begin{subequations}\label{eq:zeta1}
\begin{align}
\phi_i^\text{I}(t)  &= P_i \, \zeta_i \, \bar{\phi}_i(t) \, ,\\
\phi_i^\text{II}(t) &=  \zeta_i \, \bar{\phi}_i(t) \, ,
\end{align}
where 
\be\label{eq:partition_degree}
\zeta_{i}(P_{i},V^{\text{I}})=\frac{1}{1+(P_{i}-1)\frac{V^{\text{I}}}{V^{}}} \, 
\ee
\end{subequations}
is the partitioning degree.
The volume fractions of the clients in the two phases can be expressed in terms of the average client volume fraction as
\be
\label{eq:avg_cli_pheq}
\bar{\phi}_i(t)=\frac{1}{V}\Bigg(V^\I\phi^{\I}_{i}(t)+V^\II\phi^{\II}_{i}(t)\Bigg)\qc
\ee
where $V^\III$ denotes the respective phase volumes, and $V=V^\I+V^\II$ is the total volume. 
At phase equilibrium, the phase volume $V^\I=V (\bar{\phi}_1-\phi_1^\text{II})/(\phi_1^\text{I}-\phi_1^\text{II})$ is constant and set by the average scaffold volume fraction, $\bar{\phi}_1$. Thus, for the client dynamics, 
$V^\I$ is a control parameter. 

The chemical reactions among clients change the average volume fractions of clients in time:
\be\label{eq:avgclient}
    \frac{d}{dt}\bar{\phi}_{i} (t) = \bar{s}_{i}(\{\bar{\phi}_{j}(t)\}) \, ,
\ee
where $\bar{s}_{i}(t)=\left( V^\text{I} s^\text{I}_i + V^\text{II} s^\text{II}_i \right) / V$ are the average chemical reaction rates with the phase-dependent reaction rates $s^\III_i$   given in   Eq.~\eqref{eq:prod_rate_with_H}.
Note that $\phi_i^\III$ in $s^\III_i$  can be substituted by 
the average volume fractions $\{\bar{\phi}_{j}(t)\}$ using Eqs.~\eqref{eq:zeta1}.
Thus the average reaction rates $\bar{s}_i$ solely depend on the average volume fractions
$\{\bar{\phi}_{j}(t)\}$.

Equation~\eqref{eq:avgclient} governs the chemical reaction of client $i$ at phase equilibrium. 
Compared to the thin interface model (Sect.~\ref{sect:thin_interface}), the chemical kinetics is governed by ordinary differential equations, and the diffusion coefficients no longer determine the dynamics of the client volume fractions.   
The remaining parameters at phase equilibrium are the reaction rate coefficients  $k^{\III}_{\alpha}$, the fuel energies $\tilde{\mu}^{\III}_{F}$, the partition coefficients $P_{i}$, and the phase volumes $V^{\III}$.
Note that Eq.~\eqref{eq:avgclient} can describe chemical kinetics maintained away from chemical equilibrium by fuel energy $\tilde{\mu}^{\III}_{F}$.

\section{Application of theoretical framework}
\label{sec:App_chem}

In the following two sections, we apply the theoretical framework discussed in Sect.~\ref{model} to reacting diluted clients in two coexisting scaffold-rich and poor phases to two different types of chemical reactions (see Fig.~\ref{fig_1}(b) for an overview). 
We determine the quantities relevant for each chemical kinetics, such as stationary yields and initial rates. 
We compare such quantities with and without condensates and distinguish systems that are maintained away from chemical equilibrium by the fuel energy $\tilde{\mu}^{\III}_{F}$
and that can relax toward chemical equilibrium ($\tilde{\mu}^{\III}_{F}=0$). 

In Sect.~\ref{sect:Reaction1}, we ask how coexisting phases alter the stationary states of reversible chemical reactions, while in Sect.~\ref{sect:Reaction2}, we analyze the effects of coexisting phases on initial rates of irreversible assembly processes.
To illustrate the effects of coexisting phases on chemical reactions, we focus on a single scaffold-rich condensate phase of radius $R$ that is located in the center of a finite spherical symmetric container of radius $L$ (Fig.~\ref{fig_2}(a)).

\begin{figure*}[t]
\centering
\includegraphics[width=1.0\textwidth]{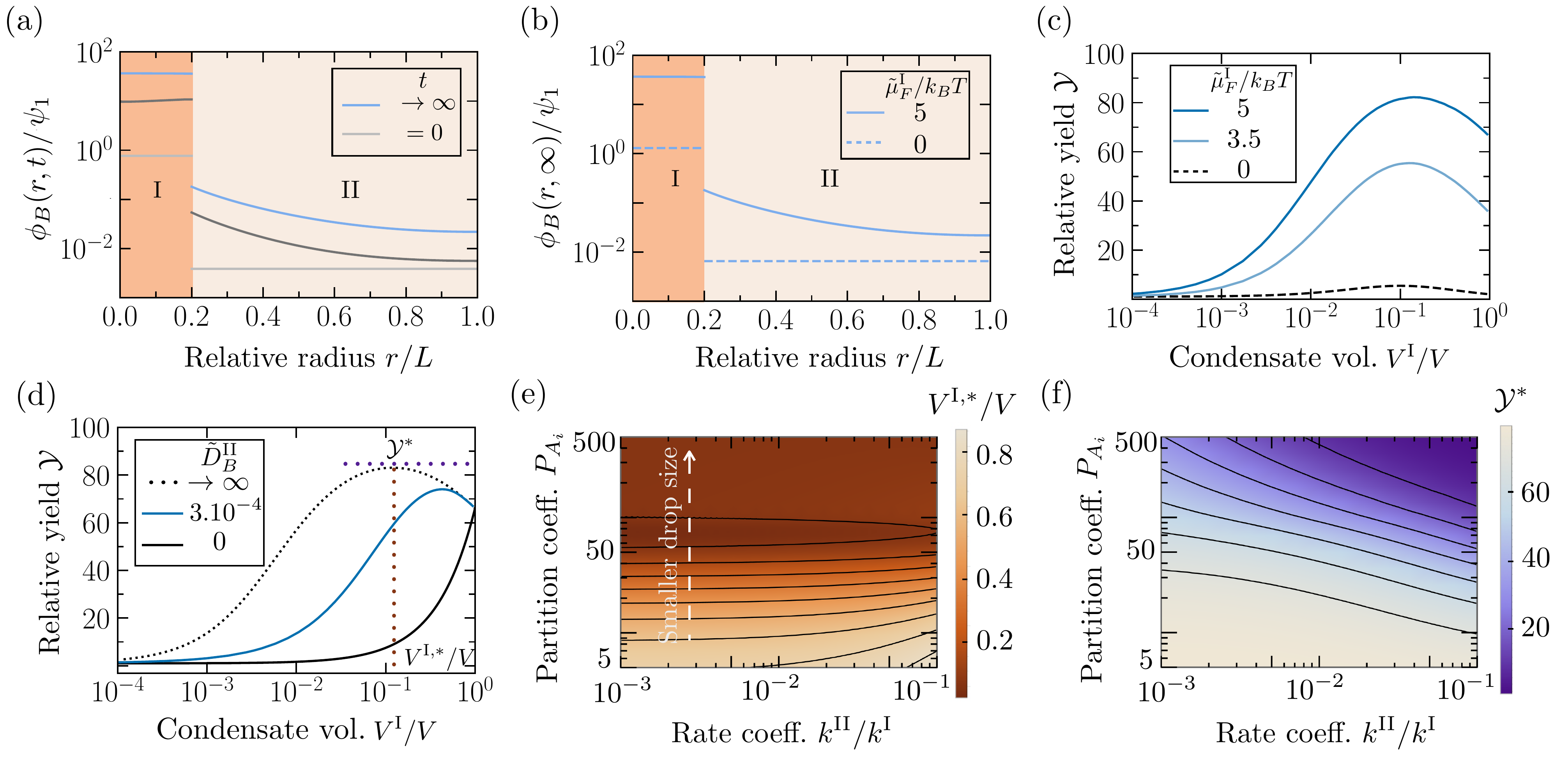}
\caption{\textbf{Condensates strongly affect the steady state of bi-molecular reactions ($A_1+A_2\rightleftharpoons B$).}
\textbf{(a)} Spatiotemporal profiles of the product component, $\phi_B(r,t)$, relative to the conserved quantity $\psi_1=(\bar{\phi}_{A_1}+\bar{\phi}_{A_2}+\bar{\phi}_{B})$.
The system initialized at phase equilibrium (light gray) approaches a non-equilibrium steady state (blue, $\tilde{\mu}^{\I}_{F}/k_{B}T=5$).
Dark gray depicts the profile at an intermediate time $t=1$. Time is rescaled by $t\to t \, k^\I$, where $k^\I$ is the reaction rate coefficient in phase I defined in Eq.~\eqref{eq:sB_bimol}. 
%
\textbf{(b)} The spatial volume fraction profiles of the product at steady state for finite diffusivity values. When chemical reactions are maintained away from equilibrium (solid, $\tilde{\mu}^{\I}_{F}/k_{B}T=5$), the profiles are spatially heterogeneous, while product volume fractions are at thermodynamic equilibrium (dashed blue line, $\tilde{\mu}^{\I}_{F}/k_{B}T=0$). 
\textbf{(c,d)}  The  yield $\mathcal{Y}$ relative to the case without condensate is maximal at a finite volume $V^{\I,*}$.
The maximal yield $\mathcal{Y}^{*}$ increases with increasing fuel energy supply $\tilde{\mu}^{\I}_F$ and the faster diffusion compared to the chemical reactions ($\tilde{D}_{i} \to \infty$). 
\textbf{(e,f)} The maximal yield $\mathcal{Y}^{*}$ increases when more substrate partitions into the condensate and the faster the reactions are in the  scaffold-rich condensate compared to the scaffold-poor phase ($k^\II/k^\I \ll 1 $).
In this limit, the corresponding condensate volume decreases.
}
\label{fig_3}
\end{figure*}

\subsection{Reversible chemical reactions controlled by a scaffold-rich condensate}\label{sect:Reaction1}

Here, we study reversible two state transitions between $g$ deactivated reactants $A_{i}$ (substrate) and the activated product $B$ with the reaction scheme 
\be\label{eq:model1}
\begin{split}
 \,\sum^{g}_{i=1} A_{i} 
\ce{<=>[\ce{\tilde{\mu}^{\III}_{F} }][\ce{ \text{spontaneous} }]}
B \, .
\end{split}
\ee
Here,  $\tilde{\mu}^{\III}_{F}$ is the fuel energy that maintains the forward reaction continuously away from chemical equilibrium, while the backward reaction can relax spontaneously toward chemical equilibrium. 
Maintaining away from chemical equilibrium is realized by considering different values of the fuel energy $\tilde{\mu}^{\III}_{F}$ 
in each of the phases.
Special cases of the reaction scheme~\eqref{eq:model1} are the uni-molecular reaction for $g=1$ and the bi-molecular reaction for $g=2$. For simplicity, we consider a single product. 
Below, we will discuss and compare the results obtained for bi-molecular reactions (Fig.~\ref{fig_3}) to uni-molecular 
(Fig.~\ref{fig_3uni}).

The bi-molecular reaction ($g=2$) has two conserved quantities, which we choose as $\psi_{1}=(\bar{\phi}_{A_1}+\bar{\phi}_{A_2}+\bar{\phi}_{B})$ and $\psi_{2}=(\bar{\phi}_{A_1}-\bar{\phi}_{A_2})$.
In our work, we consider incompressible systems and thereby restrict ourselves to volume-conserving chemical reactions, requiring that the chemical reaction rates obey: 
\begin{subequations}
\begin{align}
s^{\III}_{A_{1}}&=s^{\III}_{A_{2}}\, ,\\
s^{\III}_{A_{1}}+ s^{\III}_{A_{2}}&= - s^{\III}_{B} \, .
\end{align}
\end{subequations}
For diluted clients $A_i$ (substrates) and $B$ (product) at phase equilibrium and for $g=2$, the chemical reaction rate of the product $B$ reads
\begin{align}
\label{eq:sB_bimol}
&s^{\III}_{B}= k^{\III} \frac{\nu_B}{\nu_{A_i}}\bigg[-\exp\bigg(\frac{\bar{\mu}^0_{B}}{k_{B}T}\bigg)(\bar{\gamma}_{B}\phi_{B})^{\III} \\
\nonumber
&\quad +
\exp\bigg(\frac{\bar{\mu}^0_{A_1}+\bar{\mu}^0_{A_2}+\tilde{\mu}^{\III}_{F}}{k_{B}T}\bigg)(\bar{\gamma}_{A_1}\phi_{A_1})^{\III} (\bar{\gamma}_{A_2}\phi_{A_2})^{\III}
\bigg]\, ,
\end{align}
where molecular volumes obey $\nu_{A_i}/\nu_B=1/2$ 
for $i=1,2$ and $k^\III$ denote the reaction rate coefficients in phase I and II. 
Moreover,  $\bar{\gamma}^{\III}_{i}$ are the constant exchange activity coefficients of the clients $A_i$ and $B$, which depends on the constant scaffold volume fraction $\phi^{\III}_{1}$ in the respective phases. 
Note that at phase equilibrium (Sect.~\ref{sect:phase_eq_section}), the activity coefficients of the two phases are coupled to each other via the partitioning coefficients  $P_i=\bar{\gamma}_i^\II/\bar{\gamma}_i^\I$ ($i=A_1,A_2,B$) that are independent of client volume fraction for diluted clients.  
In the case of the thin interface model (Sect.~\ref{sect:thin_interface}), the position and time-dependent client fields $\phi_i(r,t)$  satisfy partitioning at the interface (Eq.~\eqref{eq:bc_partitioning_interface}) but not necessarily inside the phases.

Using the thin interface model for step-like and stationary scaffold profile, we numerically calculated the volume fraction profiles of the activated product $\phi_B(r,t)$. The results are shown in  Fig.~\ref{fig_3}(a,b) 
for $g=2$ and in Fig.~\ref{fig_3uni}(a,b) for $g=1$.
We could also derive analytic solutions for uni-molecular reactions ($g=1$); see Appendix~\ref{App:thin_interface_rev_uni} for details. 
Initializing the concentrations in each phase at phase equilibrium, the chemical kinetics  gives rise to spatial gradients in each phase over time. For systems that are maintained away from chemical equilibrium ($\tilde{\mu}_{F}^\III \not= 0$), gradients in the phases and, thereby, fluxes between the phases persist in the non-equilibrium steady state.

\subsubsection{Condensates and non-equilibrium driving strongly enhance the yield of reversible reactions}

To characterize the effects of phase-separated condensates on chemical kinetics, 
we consider the yield at steady state ($t\to\infty$) of the average volume fraction of the activated product, $\bar{\phi}_B(\infty)$. We define the relative yield as the ratio of the yield in the presence of a condensate of volume $V^\I$ to the yield without a condensate ($V^\I=0$):
\be \label{eq:rel_yield_def}
\mathcal{Y}=\frac{\bar{\phi}_{B}(\infty)}{\bar{\phi}_{B}(\infty)|_{V^\I=0}}\qd
\ee
The average volume fraction of each diluted client  $\bar{\phi}_i(\infty)$
at steady state is given by Eq.~\eqref{eq:avg_cli_pheq} at phase equilibrium, and for the thin interface and continuum model, it is given by $\bar{\phi}_i(\infty)=V^{-1} 4 \pi \int_0^L dr\, r^2 \phi_{i}(r,\infty)$ and $V=(4/3)\pi L^3$ (Fig.~\ref{fig_2}(a)).
In summary, the relative yield characterizes how much the compartment volume $V^\I$, which is controlled by the amount of scaffold $\bar{\phi}_1$, regulates the steady-state volume fraction of the product.

\begin{figure*}[t]
\centering
\includegraphics[width=1.0\textwidth]{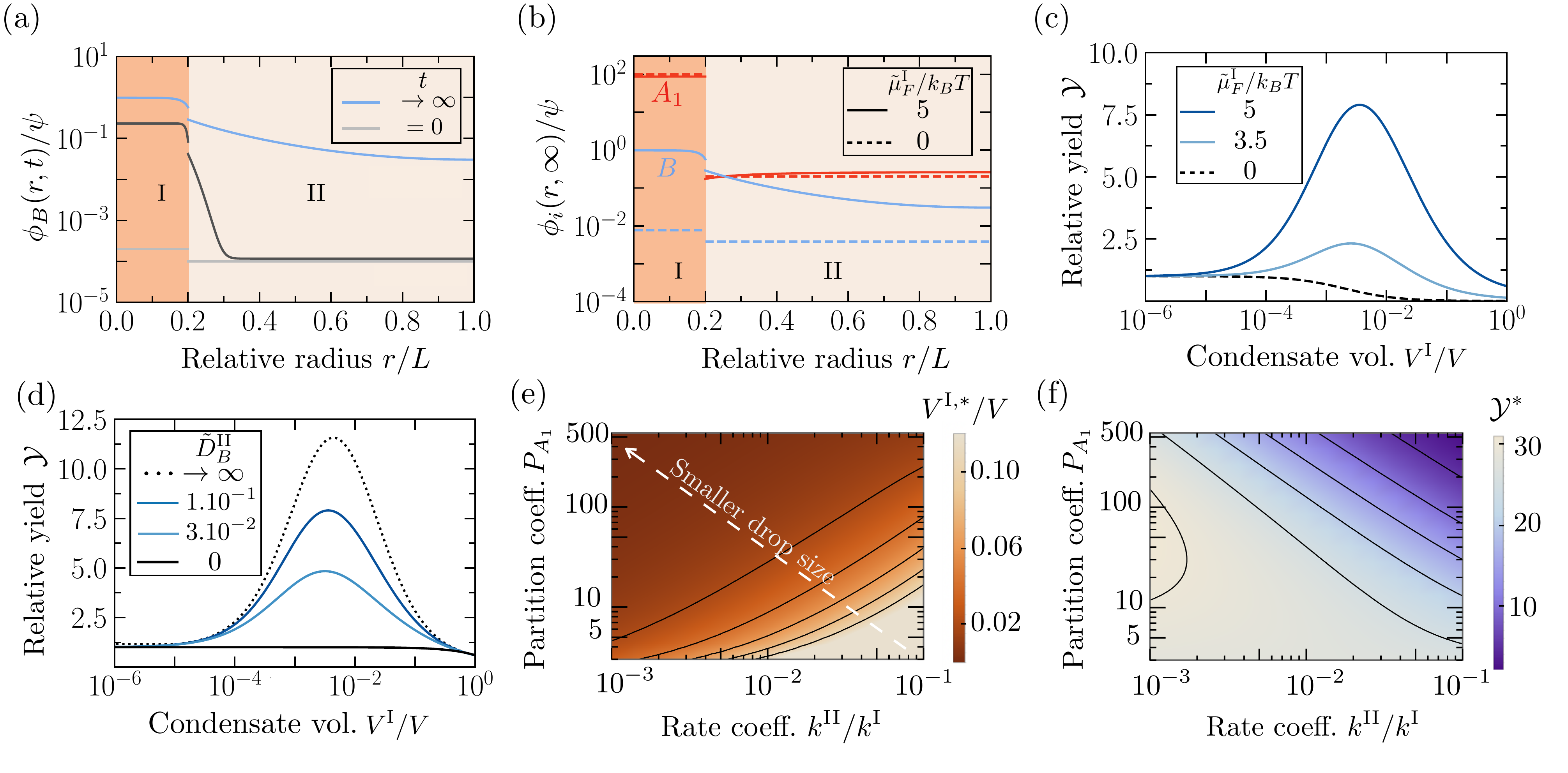}
\caption{\textbf{Condensates affect the steady-state turnover of uni-molecular reactions ($A_1 \rightleftharpoons B$).} \textbf{(a)}
Spatiotemporal profile of the product component, $\phi_B$, relative to the conserved quantity, $\psi=(\phi_{A_1}+\phi_{B})$. 
The system initialized at phase equilibrium (light gray) approaches a non-equilibrium steady state (blue, $\tilde{\mu}^{\I}_{F}/k_{B}T=5$).
Dark gray depicts the profile at an intermediate time $t=0.075$, 
where time is rescaled by $t\to t \, k^\I$ with $k^\I$ being the reaction rate coefficient in phase I.
%
\textbf{(b)} The  spatial volume fraction profiles of the clients at steady state for a finite diffusivity value, $\tilde{D}_{B}=0.01$. When chemical reactions are maintained away from equilibrium (solid, $\tilde{\mu}^{\I}_{F}/k_{B}T=5$), the profiles are spatially heterogeneous, while client volume fractions are homogeneous at thermodynamic equilibrium (dashed, $\tilde{\mu}^{\I}_{F}/k_{B}T=0$).
\textbf{(c,d)} The yield $\mathcal{Y}$ relative to the case without condensate is maximal at a finite volume $V^{\I,*}$.
The maximal yield $\mathcal{Y}^{*}$ increases with increasing fuel energy supply $\tilde{\mu}^{\I}_F$ and the faster diffusivity (black dotted) compared to the chemical reactions. 
Importantly, there is no maximum for the passive case corresponding to $\tilde{\mu}^{\I}_F=0$ for uni-molecular reactions (black dashed in (c)) and the limit of no diffusivity (black solid in (d)).
\textbf{(e,f)} The maximal yield $\mathcal{Y}^{*}$ increases when less substrate $A_1$ partitions and the larger the reaction rate coefficient in the condensate relative to the scaffold-poor phase,  $k^\II/k^\I \ll 1$.
The condensate volume corresponding to maximal yield gets smaller for decreasing  $k^\II/k^\I$ and partition coefficient $P_{A_1}$.
}
\label{fig_3uni}
\end{figure*}

We find that for the passive case ($\tilde{\mu}_{F}^\III = 0$) and the case with fuel energy supply ($\tilde{\mu}_{F}^\III \not= 0$), the relative yield $\mathcal{Y}$  can increase due to the presence of a phase-separated condensate, i.e., $\mathcal{Y}>1$; see Fig.~\ref{fig_3}(c). An exception is the uni-molecular reaction ($g=1$), where $\mathcal{Y}$ monotonously decreases with increasing condensate volume for $\tilde{\mu}_{F}^\III = 0$ (dashed line in Fig.~\ref{fig_3uni}(c)).
The increase of the relative yield depends on the value of fuel energy $\tilde{\mu}_{F}>0$. While the yield for passive systems with bi-molecular reactions increases only weakly by having a condensate (black dashed line in Fig.~\ref{fig_3}(c)),  chemical reactions maintained away from equilibrium give rise to a significantly more pronounced increase of the relative yield $\mathcal{Y}$. 
Specifically, the yield can increase by 100-fold already 
for a fuel energy supply inside the condensate $\tilde{\mu}^{\I}_{F}$ ($\tilde{\mu}^{\II}_{F}=0$) that is a  few $k_BT$.
For example, such an amount of free energy can be provided by the hydrolysis of ATP~\cite{milo:2015}.

\subsubsection{Maximal yield in reversible reactions mediated by condensates}\label{sect:max_principle_section}

A hallmark feature of our framework where clients are diluted relative to scaffold and solvent components is that the condensate volume $V^\I$ is a control parameter. 
This means that  $V^\I$ can be varied without being altered by the clients' chemical kinetics. Most importantly, we expect that the chemical kinetics of clients is affected when varying the condensate volumes 
$V^\I$. 
When increasing $V^{\I}$, we find a maximum in the relative yield (Fig.~\ref{fig_3}(c,d) and Fig.~\ref{fig_3uni}(c,d)). 
The yield maximum $\mathcal{Y}^{*}$ corresponds to a specific optimal condensate volume $V^{\I,*}$ at which $\mathcal{Y}^\prime(V^\I)|_{V^{\I,*}}=0$.  
The existence of such a maximum requires that the slope of the yield is positive at $V^\I=0$, and negative when the condensate volume is equal to the system volume, $V^\I=V$.
These conditions are sufficient for a uni-molecular reaction ($g=1$) and a bi-molecular reaction ($g=2$) since there is at most one maximum in the interval $(0,V^\I/V)$. 
The conditions for both cases are given in the Appendix, Sect.~\ref{App:4a} and \ref{App:4b}.

For passive ($\tilde{\mu}_F^\I=0$), uni-molecular ($g=1$) reactions (see Fig.~\ref{fig_3uni}(c)), we find that there cannot be any maximum in yield. The yield monotonously increases or decreases with condensate volume $V^\I$.
The reason for this behavior is related to the linearity of the uni-molecular reactions among diluted clients. 
Although partitioning of a reactant from one to the other phase can alter the chemical kinetics in each phase, there is no effect on the kinetics of the average composition and thereby on the relative yield (Eq.~\eqref{eq:rel_yield_def}). 
This is because the average compositions are proportional to the volume-weighted reaction rate coefficient, $(V^\I k^\I + V^\II k^\II)/V$, that changes monotonously with condensate volume $V^\I$.

A maximum exists when uni-molecular ($g=1$) reactions are maintained away from equilibrium ($\tilde{\mu}_F^\I \not= 0$ in Fig.~\ref{fig_3uni}(c)), or when considering bi-molecular reactions ($g=2$) (Fig.~\ref{fig_3}(c)). In both cases, a non-linear volume dependence is introduced, giving rise to two competing effects that lead to a maximum in the yield as a function of condensate volume. To understand this competition, we discuss the effects of introducing a small scaffold-rich condensate of volume $V^I/V \ll 1$. 
To obtain a maximum in the relative yield, the tiny condensate has to promote the formation of product $B$ for which we evaluate the yield $\mathcal{Y}$ (Eq.~\eqref{eq:rel_yield_def}). 
Product formation is promoted by a condensate when the reaction rate coefficients satisfy $k^\I/k^\II \gg 1$. 
Thus, increasing condensate volume also increases the product yield in the entire system. 
However, increasing condensate volume $V^\text{I}$ further toward $V^\text{I}/V=1$ leads to a decrease in the relative yield $\mathcal{Y}$  at some point. This decrease in $\mathcal{Y}$ results from a decrease in the rate of product formation $s_B$ due to a generic dilution effect inside the condensates when increasing the condensate volume $V^\I$ from a tiny volume to the volume of the system $V$. Dilution arises from a decrease in substrate volume fraction inside the condensate from $P_{A_i} \bar{\phi}_{A_i} \to \bar{\phi}_{A_i}$ for $V^\I/V\to 1$. This effect is characterized by the partitioning degree (Eq.~\eqref{eq:partition_degree}) that decreases from 1 to $P_{A_i}^{-1}$ at $V^\I/V =1$.

We further observe that the relative yield $\mathcal{Y}$  is most pronounced in the limit when client diffusion rates are fast compared to reaction rate coefficients among clients (Fig.~\ref{fig_3}(d) and Fig.~\ref{fig_3uni}(d)). This case corresponds to clients being at phase equilibrium during the reaction kinetics and is well-fulfilled when the system size is smaller than the reaction length scales (see Sect.~\ref{eq:model1} for a more detailed discussion). 
The reason why the yield is largest at phase equilibrium can be understood by considering the opposite limit, i.e., when client reactions become fast compared to diffusion of clients (black solid line in Fig.~\ref{fig_3}(d) and Fig.~\ref{fig_3uni}(d)). In this case, there are still two phases with different reaction rate coefficients $k^\I$ and $k^\II$ because the scaffold component is phase-separated, providing distinct domains for chemical reactions. 
However, as clients diffuse slowly compared to their reactions, the bulk phases cannot follow the relative partitioning at the interface. 
Thus, introducing a condensate is of no benefit except for the differences in the reaction rate coefficients. This can only lead to a monotonous increase in yield with condensate volume (for $k^\I > k^\II$). Consistently, yields become equal in both limits
$V^\I/V=0,1$  since there is no partitioning when there is no condensate or when it occupies the full system volume $V$.
Our studies show that tiny condensates are optimal bioreactors. 
The volume corresponding to maximal yield, $V^{\text{I},*}$ decreases and value of the maximal yield $\mathcal{Y}$ increases when condensates favor more the formation of the product ($k^\II/k^\I \to 0$ in Fig.~\ref{fig_3}(e,f) and \ref{fig_3uni}(e,f)). 
We also find that the larger the partitioning coefficients of the substrate $A_i$, the smaller 
the volume at maximal yield, $V^{\text{I},*}$.
Both trends support the idea that biomolecular condensates in living cells, which are usually much smaller than the system volume, can have significant effects on the yield of reversible chemical reactions that are maintained away from chemical equilibrium.

\subsection{Assembly processes controlled by a scaffold-rich condensate}\label{sect:Reaction2}
\begin{figure*}[tb]
    \centering
    \includegraphics[width=1.0\textwidth]{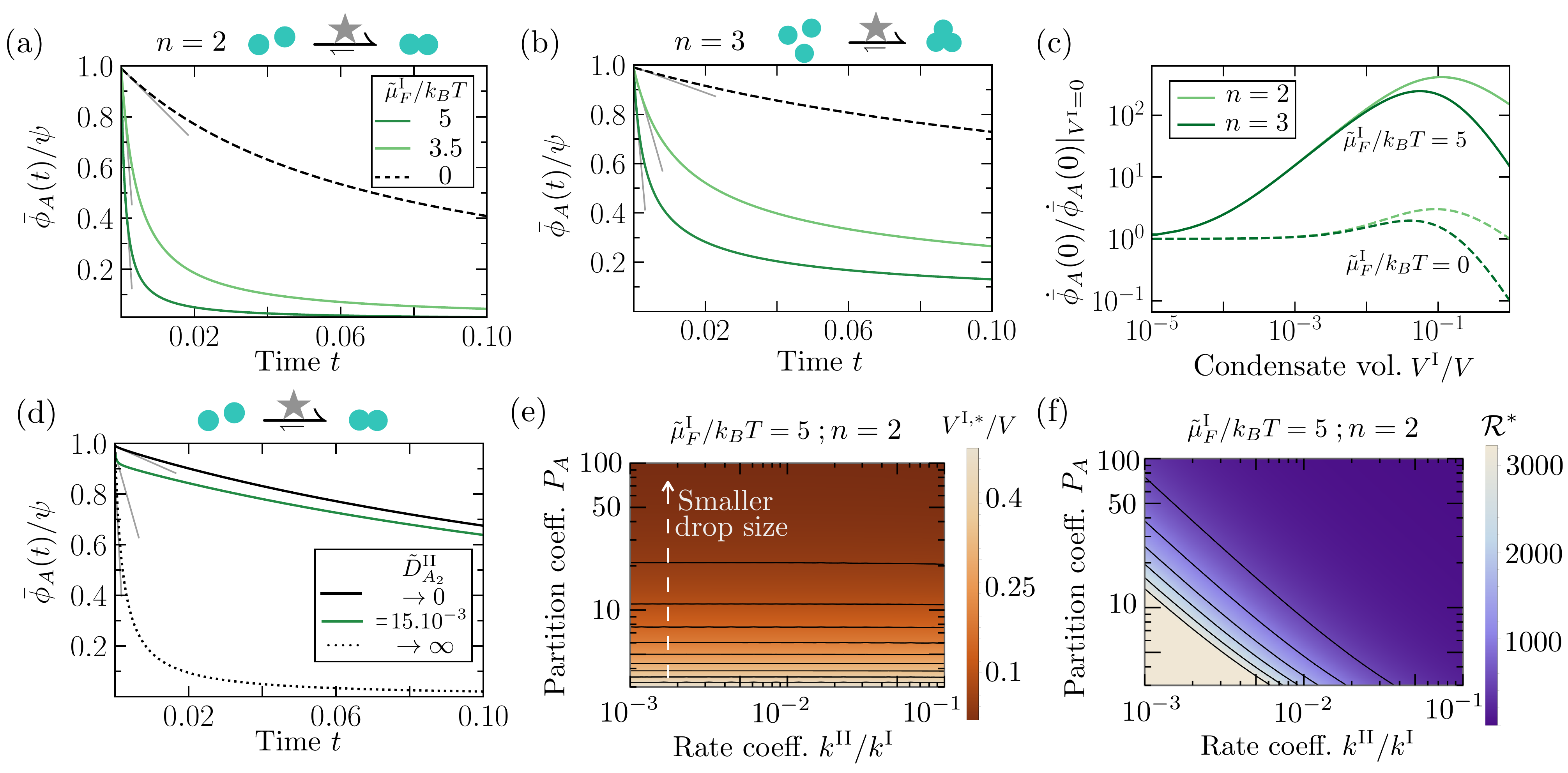}
    \caption{\textbf{Condensates can speed up assembly processes.} 
    The temporal evolution of the average volume fraction of the monomer $A$ is shown for different assembly reaction orders, \textbf{(a)} $n=2$ and \textbf{(b)} $n=3$, for passive systems ($\tilde{\mu}_F^\I=0$) and assembly processes promoted by a fuel energy supply ($\tilde{\mu}_F^\I\not= 0$) (dashed and solid lines respectively). 
    Gray lines indicate the initial decay of components $A$, $\bar{\phi}_A(t)= t \, \dot{\bar{\phi}}_A(t=0)$. The initial rate  $\dot{\bar{\phi}}_A(t=0)$ is analyzed in more detail in the following figure panels. For both plots a and b, $V^\I=0.008\,V$. 
    Time is rescaled by $t\to t \, k^\I$, where $k^\I$ is the reaction rate coefficient in phase I defined in Eq.~\eqref{eq:b2}.
    \textbf{(c)} The relative initial rate of assembly $\mathcal{R}$ (Eq.~\eqref{eq:relativeinitrate}) exhibits a maximum with condensate volume $V^\I$, for $n=2$ and $n=3$. The maximum results from a competition between the condensate-mediated promotion of assembly and a dilution effect for very large condensate volumes; see Sect.~\ref{sect:max_principle_section} for details.  
\textbf{(d)}   Acceleration of the assembly process is most pronounced at phase equilibrium (dotted line) where diffusion rates are fast compared to reaction rate coefficients. This is evident in the slope of $\bar{\phi}_A(t)$ at early times $t$.
    \textbf{(e)} The  condensate volume corresponding to maximal assembly rates decreases for larger partitioning of the monomer, $P_A$,
    while the ratio of reaction rate coefficients, $k^\II/k^\I$, has almost no impact.
    The assembly reaction order in (e) and (f) is $n=2$. 
  \textbf{(f)} The maximum amplification in assembly rates increases with decreasing  ratio of reaction rate coefficients, $k^\II/k^\I$.
 Partitioning of monomers in the condensate phase doesn't necessarily imply a higher value of $\mathcal{R}$. 
    }
    \label{fig_4}
\end{figure*}

In this section, we use the framework developed in Sec.~\ref{sec:App_chem} to study a simple model for the assembly of monomers to a nucleus composed of many monomers.  
Such a process is abundant in living cells. 
Examples are the polymerization of biofilaments and the aberrant aggregation of misfolded proteins. Condensates were suggested to control the formation of such nuclei. 
Here, we study how the initial assembly rate of nucleus formation is affected by a condensate of volume $V^\I$. 
We distinguish between a reversible assembly process that can relax towards thermodynamic equilibrium and that is maintained away by the fuel energy $\tilde{\mu}_F$. We choose the internal energy of the nucleating components higher such that the backward pathway of the assembly process is suppressed. In particular, at early times, this pathway is negligible. We only include this pathway to obtain a stationary state at long times that is consistent with thermodynamics, i.e., a state with a non-vanishing volume fraction of assembly-prone monomers. 

In this simple model for an assembly process, $n$ identical monomeric components $A$ reversibly form  a nucleus $A_{n}$ composed of $n$ monomers, where $n$ is the assembly reaction order. The corresponding reaction scheme is:
\be\label{eq:b1}
 n\,  A 
\ce{<=>>[\ce{\textcolor{white}{}\tilde{\mu}_{F}^\III \textcolor{white}{}}][\ce{\text{suppressed}}]}
A_{n} \qc
\ee
where $\tilde{\mu}_{F}^\III$ is the phase-dependent fuel energy maintaining the assembly process  away from equilibrium.
At phase equilibrium, the corresponding chemical reaction rates are
\be
\label{eq:b2}
\begin{split}
    s^{\III}_{A}&=- s^{\III}_{A_n}\\
    &= k^{\III} \frac{\nu_{A_n}}{\nu_A}\bigg[\exp\bigg(\frac{\bar{\mu}_{A_n}}{k_{B}T}\bigg)-\exp\bigg(\frac{n\bar{\mu}_{A}+\tilde{\mu}^{\III}_{F}}{k_{B}T}\bigg)\bigg] \, .
    \end{split}
\ee
where due to incompressibility the fraction of molecular volumes, $\nu_{A_n}/{\nu_A}$, is chosen to be equal to the assembly reaction order $n$. 
The conserved quantity of this assembly process is
$\psi=(\bar{\phi}_{A}+\bar{\phi}_{A_n})$.
Using Eq.~\eqref{eq:b2} in Eq.~\eqref{eq:reacdiff1}, we proceed to solve the system numerically. As initial condition, we use $\bar{\phi}_{A_n}(t)=0.01\psi$.

\subsubsection{Condensates and non-equilibrium driving can strongly accelerate assembly processes}

To characterize the effects of phase-separated condensates on the formation of assemblies, we consider the initial (at time $t=0$) assembly rate of the monomeric component $A$ relative to the case without condensates ($V^\text{I}=0$):
\be
\label{eq:relativeinitrate}
\mathcal{R}=\frac{\dot{\bar{\phi}}_{A}(0)}{\dot{\bar{\phi}}_{A}(0)\vert_{V^\I=0}} \, ,
\ee
where $\cdot=d/dt$ denotes time derivative.
We compare initial assembly rates for different assembly reaction orders $n=2$ and $n=3$, respectively.
We find that the initial rate of assembly increases with assembly reaction order $n$ (Fig.~\ref{fig_4}(a,b,c)).
This trend applies to the case with and without condensates.
We can calculate the dynamics of the monomer volume fraction for intermediate time scales, i.e.,  
for timescales when a significant fraction of monomers have assembled but the system has not yet reached the stationary state. 
We find that such intermediate times the monomer volume fraction decays algebraically, i.e.,   $\bar{\phi}_A(t)\propto t^{1/(1-n)}$, where the exponent of the power-law is set by the assembly coefficient (inset of Fig.~\ref{fig_4}(a,b); details see Appendix.~\ref{ap:hom_nuc}).

A key finding is that the presence of a fuel energy supply $\tilde{\mu}_F^\I$ strongly accelerates the assembly process; see Figs.~\ref{fig_4}(a,b,d).
To show this, we distinguish the case when assembly is maintained away from equilibrium ($\tilde{\mu}_F^\I \not= 0$) and when assembly occurs in a passive system  ($\tilde{\mu}_F^\I = 0$).
Since the backward pathway is suppressed, in particular at early times, 
the acceleration of assembly is a result of an effective increase of the monomer chemical potential by the fuel energy $\tilde{\mu}^\I_F$ inside the condensates. 
These effects are most pronounced when diffusion is fast compared to the assembly process (Fig.~\ref{fig_4}(d)). 

\subsubsection{Maximal acceleration of assembly processes mediated by condensates}

The acceleration of the assembly rate depends on condensate volume (Fig.~\ref{fig_4}(c)).
The acceleration relative to the case without condensate is maximal at a specific, optimal volume $V^{\I,*}$. 
At this volume, the relative increase in assembly rate can be significant compared to the case without condensates. This acceleration can be amplified even further to more than a hundredfold if the system is maintained away from equilibrium ($\tilde{\mu}_F^\I > 0$). 

The condensate volume at which the assembly rate is maximal, $V^{\I,*}$, and the corresponding assembly rate $\mathcal{R}^*$ depend on the relative assembly rate constants $k^\II/k^\I$ and the partition coefficient of the monomers $P_A$ (Fig.~\ref{fig_4}(e,f)). 
We find that the optimal volume $V^{\I,*}$ is mostly determined by monomer partitioning $P_A$ and that  $V^{\I,*}$ decreases with increasing $P_A$.  
The maximal assembly rate $\mathcal{R}^{*}$ is however strongly affected by both $P_A$ and $k^{\II}/k^{\I}$. It is larger the higher the assembly rate constant inside, $k^{\I}$, compared to outside, $k^{\II}$, and the smaller the partition coefficient of monomers $P_A$.

\section{Conclusion}

Unraveling the physicochemical principles of how condensates regulate chemical processes is challenging due to the interplay between chemical reactions and phase separation. 
To dissect the effects of condensates on chemical kinetics, we developed a general theoretical framework for diluted clients undergoing chemical reactions in a phase-separated environment where scaffold and solvent components form a condensate. We derived the equations governing chemical kinetics for diluted clients and showed that they undergo chemical reactions in a spatially heterogeneous environment determined by the phase-separated scaffold component. 
This environment gives rise to an effective drift that originates from cross-diffusion. This drift drives the diffusive exchange through the condensate interface while the clients follow reaction-diffusion kinetics with phase-dependent transport coefficients.
In the thin interface limit, the drift  gives rise to distinct boundary conditions at the condensate interface. 
These boundary conditions entail the effects of how a condensate affects chemical kinetics.

We illustrate the effects of a condensate on chemical processes  by considering two examples, namely a reversible reaction between a substrate and product and an assembly process from a substrate to a product. 
We determined the product yield for the reversible reactions and initial rates for the assembly process. A key finding is that both quantities can be maximal at a specific condensate volume. 
This maximum results from a competition of the condensate promoting the formation of products and a dilution effect of the substrate inside the condensate.  
We also found that when diffusion is fast compared to chemical processes (i.e., the system is at phase equilibrium) and when maintaining chemical processes away from chemical equilibrium, the effects of condensates on chemical kinetics can be amplified significantly.

Regulating yields and initial speeds of chemical processes is key in biological systems. For example, cells have to silence expression upon stress~\cite{du2023condensate} or control the formation of biofilaments~\cite{hernandez2017local} and aberrant aggregates~\cite{patel2015liquid}. From a physical chemistry perspective, condensates have a great propensity to provide switch-like mechanisms between regimes strongly differing in the properties of chemical processes. This switch can for example be achieved by  cycles of condensate formation and dissolution. In particular, since chemical processes in cells are maintained away from equilibrium, such switches can be extraordinarily pronounced, according to our theoretical studies. 

Condensate-mediated switches may have also played an important role in the molecular origin of life. 
The non-dilute environments leading to condensed phases should have been abundant at Early Earth, as non-dilute conditions can be easily created through drying~\cite{Fares:2020, Tekin:2022, haugerud2024nonequilibrium} or freezing processes~\cite{mutschler2015freeze}. In particular, in the presence of cycles (temperature, salt, etc.),  switching the speed of chemical processes could have provided selection mechanisms for specific molecules that are based on the physical non-equilibrium conditions~\cite{Barto:2023}.

Future applications of our theoretical framework might address questions in systems with extremely many components, such as mixtures composed of DNA and RNA differing in their sequences~\cite{bartolucci2023interplay}. In such systems, most sequences are diluted enabling to describe the reaction kinetics among sequences as diluted clients. 
Moreover, our framework could be used to systematically study the Turing patterns~\cite{epstein2016reaction,halatek2018rethinking} in systems with coexisting phases~\cite{menou2023physical}.

\acknowledgments{
We are grateful for the very helpful feedback from H.\ Vuijk for the detailed discussions on various concepts discussed in the manuscript.
We are also grateful for the feedback of S.\ Gomez and G.\ Granatelli on the manuscript. 
We thank Tyler S.\ Harmon  for pointing out multi-step reactions as an interesting application of the client-scaffold model and W.\ P\"onisch for fruitful discussions on the subject. 
We are also grateful for the discussions and the collaborations on experimental realizations with D.\ Tang, A.\ Ghosh, and M.\ Gao, as well as J.\ Boekhoven.   
F.\ J\"ulicher acknowledges funding by the Volkswagen Foundation. 
T.\ C.\ T.\ Michaels thanks support from ETH Zurich and the Swiss National Science Foundation (grant $200021\_219703$).
C.\ Weber acknowledges the 
European Research Council (ERC) for financial support under the European Union’s Horizon 2020 
research and innovation programme (``Fuelled Life'' with Grant agreement No.\ 949021).
}

\appendix

\section{Diluted clients in a phase-separated system}\label{App:1}

In this section, we derive the conditions for phase equilibrium for a mixture composed of non-dilute scaffold and solvent components, and  diluted clients that undergo chemical reactions. 
Conceptionally, it is key to define the meaning of client components being diluted in a chemically reactive mixture.  

\subsection{Thermodynamics and phase equilibrium of a $(N+2)$-component mixture}
\label{sect:app_thermodynamics_NPLUS2COMPMIX}

Due to the incompressibility condition $\sum_{i=0}^{\tot+1} \phi_i =1$, 
the $(\tot+2)$-component mixture  can be described by $(\tot+2)$ volume fractions $\phi_i$, whereby $i=0$ denotes the solvent, $i=1$ denotes the scaffold, and $i=2, ..., (\tot+1)$ label the $\tot$ dilute clients components.
The solvent volume fraction can be substituted by the relationship,
$\phi_0=1-\sum_{i=1}^{\tot+1} \phi_i$, leading to  $(\tot+1)$ independent volume fractions. 
The thermodynamics of this $(\tot+2)$-component mixture is governed by the Helmholtz free energy density of the form:
\be \label{eq:free_energy_density_app}
f=f_{0}+\sum^{\tot+1}_{i,j=0}\frac{\tilde{\kappa}_{ij}}{2}\nabla\phi_{i}\cdot\nabla\phi_{j} \, , 
\ee
where $f_{0}$ is the homogeneous Helmholtz free energy density that solely depends on the volume fractions of all components, $\phi_i$. 
The term $\tilde{\kappa}_{ij}$ in Eq.~\eqref{eq:free_energy_density_app} characterizes the free energy costs due to gradients in volume fractions. 
For simplicity, we neglect cross couplings, i.e.,  $\tilde{\kappa}_{ij}=0$ for $i\neq j$, using $\tilde{\kappa}_{ij}= \delta_{ij} \kappa_i $.
Note that the free energy costs $\kappa_i$  also contribute to the chemical reaction rates $s_i$ (Eq.~\eqref{eq:prod_rate_with_H}). However, since $\kappa_i>0$, this contribution solely shifts up  the diffusion coefficient $D_{i}=M_{ii} \sum_i \partial \bar{\mu}_i/\partial \phi_i \to D_{i} + \kappa_i$, and can thus be neglected in the following. 

Phase equilibrium between two homogeneous phases $\I$ and $\II$ in the ($\tot+2$)-component mixture are governed by the balance of the exchange chemical potentials $\bar{\mu}_i^\text{\III}$ and the osmotic pressures $\Pi^{\III}$ between the phases:
\begin{subequations}
\label{eq:phase_eq}
\begin{align}\label{eq:phase_eq_a}
    \bar{\mu}^{\I}_{i}(\{\phi^{\I}_{j}\}) &=\bar{\mu}^{\II}_{i}(\{\phi^{\II}_{j}\}) \, , \\
    \label{eq:phase_eq_b}
\Pi^{\text{I}}&=\Pi^{\text{II}}+\frac{2\gamma}{R} \, , 
\end{align}
\end{subequations}
The homogeneity of phases implies that the gradient-free energy cost vanishes and thus the exchange chemical potential can be calculated via     
$\bar{\mu}_i=\nu_i \partial f_0/\partial \phi_i$, 
and the osmotic pressure is given by $\Pi=-f_0 + \sum_{i=1}^{N+1} \phi_i \bar{\mu}_i/\nu_i$.
Here,  $\nu_i$ is the molecular volume of component $i$. 
The replacement of the solvent volume fraction implies $(N+1)$ balance equations for the exchange chemical potentials $\bar{\mu}_i$ with $i=1,..., (N+1)$ (Eq.~\eqref{eq:phase_eq_a}).
Moreover, the exchange chemical potential measures the chemical potentials relative to the solvent component, $i=0$. 
The volume of phase I is set by the average volume fractions $\bar{\phi}_i$ by the relationship 
$V^{\text{I}}/V= (\avgvf_{i}-\phi^{\II}_{i})/(\phi^{\I}_{i}-\phi^{\II}_{i})$, with $V^\II/V=1-V^\I/V$, where $V$ is the system volume.

To illustrate the implications when $\tot$ reacting clients are diluted, we consider the following mean-field, homogeneous free energy density of the form:
\be
\label{eq:Ap_freeenergy}
f_{0}=\frac{k_{B}T}{\nu_0}\bigg[\sum^{\tot+1}_{i=0}\frac{\phi_{i}}{r_{i}}\log(\phi_{i})+\sum^{\tot+1}_{i,j=0}\frac{\chi_{ij}}{2}\phi_{i}\phi_{j}+\sum^{\tot+1}_{i=0}\omega_{i}\phi_{i}\bigg] \, , 
  \ee
  where $r_i=\nu_{i}/\nu_0$ is the fraction of molecular volumes $\nu_i$.
Moreover, the first term represents the entropic contribution of all components. The second term describes the mutual interactions among all components with pair-wise interaction parameter, $\chi_{ij}$.
The homogeneous free energy density $f_0$ does not depend on the solvent volume fraction $\phi_0$ due to $\phi_0=1-\sum_{i=1}^{\tot+1} \phi_i$. Note that $\phi_0$ has not been explicitly replaced in Eq.~\eqref{eq:Ap_freeenergy} for presentation purposes. We remark that our  framework for diluted and reacting clients is also valid for free energies that take into account interactions beyond mean-field; the mean-field free energy above is only chosen for simplicity.

Using Eq.~\eqref{eq:Ap_freeenergy}, the exchange chemical potentials read 
\begin{align}
\nu_i \partial f_0/\partial \phi_i &= k_{B}T\left[1-r_i+r_{i}(\omega_i-\omega_0+\chi_{0i})\right] \\
\nonumber
&\quad +k_{B}T\log(\bar{\gamma}_{i}\phi_{i}) 
\qd
\end{align}
 Moreover, the exchange activity coefficients have the following form: 
\be \label{eq:activity_coeff_app}
\bar{\gamma}_{i} = \frac{1}{(1-\sum^{\tot+1}_{j=1}\phi_{j})^{r_i}}\exp\left[\sum^{\tot+1}_{j=1}r_{i}(\chi_{ij}-\chi_{0i}-\chi_{0j})\phi_{j}\right]\qd
\ee
Comparing the two relationships above with the general form of the exchange chemical potential Eq.~\eqref{eq:chem_pot},
noting that the exchange chemical potential for spatially heterogeneous system reads $\bar{\mu}_{i}=\nu_i \partial f_0/\partial \phi_i -\kappa_i \nabla^2 \phi_i$, 
the reference chemical potentials are $\bar{\mu}_{i}^0=k_{B}T[1-{r_i}+r_{i}(\omega_i-\omega_0+\chi_{0i})]$.

For an ideal solution and components having a non-zero molecular volume, $\chi_{ij}=0$, and thus, the exchange activity coefficient  $\bar{\gamma}_{i} = (1-\sum^{\tot+1}_{j=1}\phi_{j})^{-r_i}$.
Note that only for ideal mixtures composed of point particles where all volume fractions except the one of the solvent vanish to zero, $\bar{\gamma}_{i} =1$.

\subsection{Thermodynamics of $N$ diluted clients}

In this section, we derive  approximate  conditions for phase equilibrium (Eq.~\eqref{eq:phase_eq}) and approximate expressions for the exchange activity coefficient (Eq.~\eqref{eq:activity_coeff_app}) 
when  the $N$ reacting clients ($i=2,..., N+1$)  are diluted compared to scaffold component ($i=1$) and solvent  ($i=0$). Diluted means that the client volume fractions
$\phi_i$ are much smaller than the volume  fractions of scaffold and solvent, $\phi_1$ and $\phi_0$, respectively:
\begin{equation}
   \phi_i \ll \phi_0\, , \phi_1 \qc \quad  i=2,..., (N+1) \qd
\end{equation}
 
\subsubsection{Limit of vanishing client volume fractions}\label{app:zeroclient}

We first discuss the limit of zero volume fraction of the client components, i.e., $\phi_i/\phi_1 \to 0$ for $i=2,..., (N+1)$, and where the volume fraction of the scaffold approaches $\phi_1 \to (1-\phi_0)$. 
From Eq.~\eqref{eq:activity_coeff_app}, the client exchange activities for vanishing client volume fractions become:
\be \label{eq:activity_coeff_app_zero}
\bar{\gamma}_{i}|_{\{\phi_j=0\}} = 
\frac{1}{(1-\phi_{1})^{r_i}}
\exp\left[r_{i}(\chi_{1i}-\chi_{0i}-\chi_{01})\phi_{1}\right]\qc
\ee
where $\{\phi_j=0\}$ is an abbreviation for all client volume fractions being zero,  $\{\phi_2=0, ..., \phi_{N+1}=0\}$. 
Moreover, $r_i=\nu_i/\nu_0$, where $\nu_i$ are molecular volume of component $i$.

At phase equilibrium (Eq.~\eqref{eq:phase_eq_a}), the definition of the partitioning coefficients, $P_i\equiv\phi_i^\I/\phi_i^\II$, can be expressed in terms of the exchange activity coefficients,   $P_{i}=\bar{\gamma}_{i}^\II/\bar{\gamma}_{i}^\I$~\cite{BauermannLaha:2022}. 
For vanishing client volume fraction, the partitioning coefficients solely depend on the scaffold volume fraction $\phi_1^\III$ and read:
\begin{align}
\label{eq:App_partition_zero}
    P_{i}|_{\{\phi_j^\III=0\}} &=\bigg(\frac{1-\phi^{\I,0}_{1}}{1-\phi^{\II,0}_{1}}\bigg)^{r_i} \\
    \nonumber
    &  \times
    \exp[r_{i}(\phi^{\I,0}_{1}-\phi^{\II,0}_{1})(\chi_{01}+\chi_{0i}-\chi_{1i})] \qd
\end{align}
This expression can be rewritten using the phase equilibrium condition for the scaffold component ($\bar{\mu}_1^\I=\bar{\mu}_1^\II$): 
\begin{align}
    P_{i}|_{\{\phi_j^\III=0\}} &=   
    \left( \frac{1-\phi_1^\I}{1-\phi_1^\II}\right)^{r_i-r_i/r_1}\\
    \nonumber
    & \times
    \exp[r_{i}(\phi^{\I,0}_{1}-\phi^{\II,0}_{1})(\chi_{0i}-\chi_{1i})+1 - r_1^{-1}] \qd
\end{align}
For equal molecular volumes of scaffold and solvent ($\nu_0=\nu_1$, implying $r_1=1$, the partitioning coefficients in the dilute limit were derived recently~\cite{weber2019spatial, ponisch2023aggregation}: 
\begin{equation}
    P_{i}|_{\{\phi_j^\III=0\}} =  \exp[r_{i}(\phi^{\I,0}_{1}-\phi^{\II,0}_{1})(\chi_{0i}-\chi_{1i})] \qd
\end{equation}
In summary, the exchange activity coefficients and the partitioning coefficients each approach constant values for $\phi_i\to 0$ ($i=2,...,N+1$) that are independent of the client volume fractions $\{\phi_i\}$.

\subsubsection{Phase equilibrium conditions for small client volume fractions: Systematic expansion}

Before we perform a systematic expansion, we shortly review  the  physical and mathematical reasoning of what equations should be expanded. 
A first choice could be the dynamic equations governing diffusive transport and chemical reactions (Eq.~\eqref{eq:kinetic_eq}). Expanding such equations is always possible around any set of volume fractions. Expanding up to the $n$-th order, one obtains a set of coupled partial differential equations of order $n$ in the volume fractions. 
In particular, an expansion up to the first of the dynamic equations would also linearize chemical reactions.
The validity of this linearization is restricted to spatial variations of client volume fractions that are small deviations from the expansion point. Larger deviation from the expansion point may lead to nonphysical results such as negative volume fractions. 
Another reason against the expansion of the dynamic equations is that the effects of   clients being diluted should already be accessible on the level of the thermodynamic conditions. Note that the conditions for phase equilibrium govern  the  boundary conditions in the thin interface model (Sect.~\ref{sect:thin_interface}) and the dynamics of the average volume fractions when diffusion is fast compared to chemical rates (Sect.~\ref{sect:phase_eq_section}). 

An alternative could be the expansion of the chemical potentials since their spatial gradients  drive diffusive fluxes and their differences between chemical states drive chemical reactions. 
However, once the expansion point in volume fractions approaches zero, each chemical potential diverges due to the logarithmic dependence in $\phi_i$ (Eq.~\eqref{eq:chem_pot}). 
In other words, an expansion of the chemical potentials around zero volume fraction is not defined. 
Expanding only the non-diverging interaction terms remains conceptionally unsatisfactory~\cite{weber2019spatial}.

Another possibility is to expand the phase equilibrium conditions around zero client volume fractions. 
At phase equilibrium, the equality of chemical potentials cancels the divergence. 
Using Eq.~\eqref{eq:chem_pot} and the equilibrium conditions~\eqref{eq:phase_eq_a}, one gets $\log (\phi_i^\I/\phi_i^\II)= \log(\bar{\gamma}_i^\II/\bar{\gamma}_i^\I)$, which has no divergence for vanishing average client volume fraction ($\bar{\phi}_i \to 0$) since $\phi_i^\I \sim\bar{\phi}_i$ and $\phi_i^\II\sim\bar{\phi}_i$. In particular, for the client volume fractions $\phi_i \to 0$ with $i=2,..., (N+1)$, 
$\bar{\gamma}_i^\II/\bar{\gamma}_i^\I$
become the constant partition coefficients $P_{i}|_{\{\avgvf_j=0\}}$ (Eq.~\eqref{eq:App_partition_zero}) that do not depend on  
the client volume fractions.

In summary, expanding the equilibrium conditions~\eqref{eq:phase_eq} or, equivalently, the partitioning coefficients $P_i(\phi_1,\phi_2, ...,\phi_{N+1})$ for small average client volume fractions $\bar{\phi}_i$ appears as the most general procedure. 
For clients $i=2,..., (N+1)$ with $\phi_i \to 0$, not only the partition coefficient approaches a constant, but also the exchange activity coefficients become constants in this limit (Eq.~\eqref{eq:activity_coeff_app_zero}). Thus, equivalently, 
exchange activity coefficients can be expanded systematically.

Now we continue with the systematic expansion by expressing the equilibrium volume fractions in each phase $\phi^\III_{i}$ and the phase volume $V^\I$ as a function of the average volume fractions for the scaffold $\avgvf_{1}$, and the clients $\avgvf_{i}$ (Eq.~\eqref{eq:avg_cli_pheq}):
\be
\begin{split}
&\phi^\III_{i}=h^\III_{i}(\avgvf_{1},\avgvf_{2},..,\avgvf_{N+1}) \qc\\
&\frac{V^\I}{V}= g(\avgvf_{1},\avgvf_{2},..,\avgvf_{N+1}) \qd
\end{split}
\ee
In the following, we  expand the client average volume fraction $\{\avgvf_{i}\}$ around zero average volume fraction $\{\avgvf_{i}=0\}$ ($i=2,..,\tot+1$). Consistently, the zeroth order of this expansion will lead to the results discussed in Sect.~\ref{app:zeroclient}.

\begin{figure*}[tb]
\includegraphics[width=1.0\textwidth]{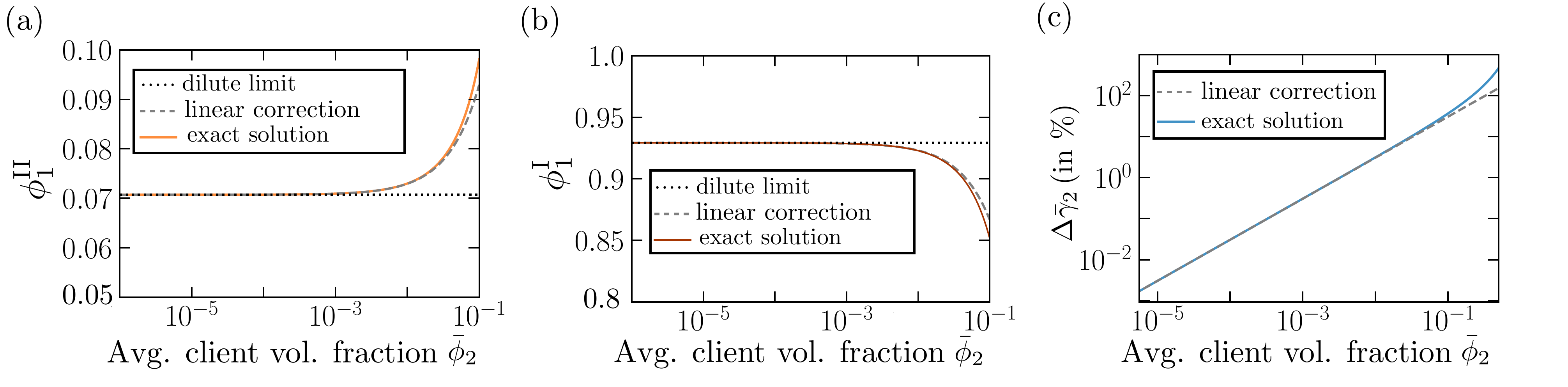} \caption{\textbf{Deviations to dilute limit of clients.} 
We quantify the effects of a non-diluted client  component ($i=2$) on the equilibrium volume fractions of scaffold $\phi^\III_1$. %
\textbf{(a,b)} The zeroth order $\phi_1^{\III,0}$ (dilute limit) is shown by the horizontal dotted line, while the dashed line is the linear order correction $\phi_1^{\III,1}$.
The orange line corresponds to the full numerical solution to Eq.~\eqref{eq:phase_eq}, indicating a good agreement of the linear order correction up to volume fraction around 0.1. For the shown curves, the zeroth order becomes a good approximation for a client volume fraction below 0.01. 
\textbf{(c)} 
We show the relative difference in the client's exchange activity coefficient, $ \Delta \bar{\gamma}_2 $ (Eq.~\eqref{eq:delta_gamma}), as a function of client volume fraction ${\phi}_2$.
The dashed line shows the relative linear correction, ${\phi}_2 / ({\phi}_2^* \bar{\gamma}^0_2)$
(Eq.~\eqref{eq:gamma1}), and the blue solid line results from  numerically solving Eq.~\eqref{eq:phase_eq}.
The parameters are: $\chi_{01}=3$, $\chi_{02}=-1$, $\chi_{12}=0$, $\phi^{0}_{1}=0.05$. The volume fraction scale (Eq.~\eqref{eq:vol_scale_2}) obtained for these parameters is ${\phi}_{2}^{*}=0.344$ and $\bar{\gamma}^{0}_{2}=0.95246$. $\Delta \bar{\gamma}_{2}=1\%$ for $\bar{\phi}_{2}=3.27\cdot10^{-3}$.
}    \label{fig_9}
\end{figure*}

In linear order, the scaffold equilibrium volume fraction can be written as
\begin{subequations}
\be
\label{eq:scaffold_expansion}
\phi^{\III}_{1} 
= \phi^{\III,0}_{1} + \sum^{\tot+1}_{i=2} \alpha^\III_{i}\avgvf_{i} + \mathcal{O}(\avgvf_{i})^2 \qc
\ee
where we abbreviated the zeroth order $\phi^{\III,0}_{1}$, and 
the linear expansion coefficients are defined as $\alpha^\III_{i}={\partial h^{\III}_{1}}/{\partial\avgvf_{i}}|_{\{\avgvf_{j}=0\}}$.

The coefficients 
$\alpha^\III_{i}$ are derived by linearizing both sides of Eq.~\eqref{eq:phase_eq_b}.
Moreover, we use the chemical potential balance (Eq.~\eqref{eq:phase_eq_a}), and  express changes of the scaffold volume fractions $\phi^{\III}_{1}$ with client volume fractions $\phi^{\III}_{i}$  in terms of $\bar{\phi}_{i}$ by utilizing the zeroth order partition coefficients $P^{0}_{i} \equiv P_{i}|_{\{\avgvf_j=0\}}$ ($i=2,...,N+1$)~\cite{Lahathe:2023}:
\be
\label{eq:app_slopes1}
\begin{split}
& \alpha^{\I}_{i}(\avgvf^{0}_{1},\{\avgvf_{i}\}=0) = a^{\I}_{i} \frac{P^{0}_{i}}{1+(P^{0}_{i}-1)V^{\I,0}/V} \qc \\
& \alpha^{\II}_{i}(\avgvf^{0}_{1},\{\avgvf_{i}\}=0) = a^{\II}_{i}\frac{1}{1+(P^{0}_{i}-1)V^{\I,0}/V} \qc \\
\end{split}
\ee
where the coefficients 
\begin{align}
a^{\text{I}}_{i}&=-\phi^{\I,0}_{1}\,
 \frac{ 1+\left(\Delta \chi_i+\frac{1-1/P^{0}_{i}}{r_{i}(\phi^{\I,0}_{1}-\phi^{\II,0}_{1})}\right)\left(1-\phi^{\I,0}_{1}\right)}{\frac{\left(1-\phi^{\I,0}_{1}\right)}{r_{1}}+\phi^{\I,0}_{1}-2\chi_{01}\phi^{\I,0}_{1}\left(1-\phi^{\I,0}_{1}\right)} \qc
 \\
 a^{\text{II}}_{i}&=-\phi^{\II,0}_{1}\,\frac{1+\left(\Delta \chi_i +\frac{P^{0}_{i}-1}{r_{i}(\phi^{\I,0}_{1}-\phi^{\II,0}_{1})}\right)\left(1-\phi^{\II,0}_{1}\right)}{\frac{\left(1-\phi^{\II,0}_{1}\right)}{r_{1}}+\phi^{\II,0}_{1}-2\chi_{01}\phi^{\II,0}_{1}\left(1-\phi^{\II,0}_{1}\right)} \qd
\end{align}
\end{subequations}
In the equations above,  we abbreviate \begin{equation}
\label{eq:abb_delta_chi}
    \Delta \chi_i = \chi_{1i}-\chi_{0i}-\chi_{01} \, .
\end{equation}
Fig.~\ref{fig_9}(a,b) shows how the volume fraction of the scaffold component in each phase, $\phi_1^\III$, behaves for increasing average client volume fraction $\bar{\phi}_2$. The horizontal dotted line is the zeroth order $\phi_1^{\III,0}$, the dashed line the linear order $\phi_1^{\III,1}$, and the orange line is the full numerical solution to Eq.~\eqref{eq:phase_eq}.

The client equilibrium volume fractions ($i=2,..., N+1$) can be expanded as follows:
\begin{align}
\label{eq:client_expansion}
\phi^{\III}_{i} 
&=\sum^{\tot+1}_{j=2}\left.\frac{\partial h^{\III}_{i}}{\partial \avgvf_{j}}\right|_{\{\avgvf_{l}=0\}}\avgvf_{j}  
\\
\nonumber
& \quad
+\frac{1}{2}\sum^{\tot+1}_{j,k=2}\left.\frac{\partial^{2}h^{\III}_{i}}{\partial \avgvf_{j}\partial \avgvf_{k}}\right|_{\substack{\{\avgvf_{l}=0\} \\l\neq i}}\avgvf_{j}\avgvf_{k}+\mathcal{O}(\avgvf_{i})^3 \qc 
\end{align}
where the zeroth order vanishes, i.e., 
$h^{\III}_{i}(\avgvf^{0}_{1},\{\avgvf_{i}=0\})$.
Thus, we introduce the fraction ${\phi^{\III}_{i}}/{\avgvf_{i}}$ that is finite when evaluating the limit of zero client volume fractions: 
\begin{align}
\frac{\phi^{\III}_{i}}{\avgvf_{i}} &= \frac{\partial h^\III_{i}}{\partial \avgvf_{i}} + \frac{1}{2}\sum^{\tot+1}_{j=2}\left.\frac{\partial^{2}h^\III_{i}}{\partial \avgvf_{i}\partial \avgvf_{j}}\right|_{\{\avgvf_{l}\}=0} \avgvf_{j}+\mathcal{O}(\avgvf_{i})^2 \qd
\end{align}
We see that the zeroth order in the expansion in client volume fractions  gives a linear order leading term. In other words, there is an order shift due to exclusively expanding around zero client volume fractions. Thus, the $n-th$ order for the scaffold expansion (Eq.~\eqref{eq:scaffold_expansion}) corresponds to a $(n+1)-th$ order in the client expansion.

The expansion of the condensate phase volume  $V^\I$ can be written as:
\begin{subequations}
\begin{align}
\nonumber
\frac{V^{\I}}{V} &= \frac{V^{\I,0}}{V} +\sum^{\tot+1}_{i=2} \left.\frac{\partial g}{\partial \avgvf_{i}}\right|_{\{\avgvf_{i}\}=0}\avgvf_{i} +\mathcal{O}(\avgvf_{i})^2\\
& = \frac{\avgvf^{0}_{1}-\phi^{\II,0}_{1}}{\phi^{\I,0}_{1}-\phi^{\II,0}_{1}} + \sum^{\tot+1}_{i=2} \eta_{i}\avgvf_{i} +\mathcal{O}(\avgvf_{i})^2 \qc
\end{align}
where the coefficients 
\begin{align}
    \eta_i &= -\frac{(\bar{\phi}^{0}_{1}-\phi^{\II,0}_{1})}{(\phi^{\I,0}_{1}-\phi^{\II,0}_{1})^2}\alpha^{\I}_{i}
    \\
    \nonumber
    &\quad +\bigg[\frac{(\bar{\phi}^{0}_{1}-\phi^{\II,0}_{1})}{(\phi^{\I,0}_{1}-\phi^{\II,0}_{1})^2}-\frac{1}{(\phi^{\I,0}_{1}-\phi^{\II,0}_{1})}\bigg]\alpha^{\II}_{i} \qd
\end{align}
\end{subequations}
The expressions of $\alpha^{\III}_{i}$ are obtained from Eq.~\eqref{eq:app_slopes1}.

The exchange activity coefficient of all components can be expanded up to the first order around $\{\phi_{i}=0\}$: 
\begin{subequations}
\begin{align}
\bar{\gamma}_{1} &=
\frac{1}{(1-{\phi}^{0}_{1})^{r_1}}\exp{(-2r_1\chi_{01}{\phi}^{0}_{1})} 
\\
\nonumber
&+ \sum^{N+1}_{i=2} r_1\frac{\exp{(-2r_1\chi_{01}{\phi}^{0}_{1})}}{(1-{\phi}^{0}_{1})^{r_1+1}}\left(1+(1-{\phi}^{0}_{1})
\Delta \chi_i\right){\phi}^{}_{i} 
\\
&+\mathcal{O}(\phi_{i}^2)
\nonumber
\qc\\
\label{eq:first_order_act_coeff}
\bar{\gamma}_{i}  
&=\frac{1}{(1-{\phi}^{0}_{1})^{r_i}}\exp{(r_i{\phi}^{0}_{1}\Delta \chi_i)}\\
\nonumber
&+\sum^{N+1}_{j=2}r_j\frac{\exp{(r_j{\phi}^{0}_{1}\Delta \chi_j)}}{(1-{\phi}^{0}_{1})^{r_j+1}}\left(1-2(1-{\phi}^{0}_{1})\chi_{0j}\right){\phi}_{j} \\
&+\mathcal{O}(\phi_{j}^2)
\nonumber
\qc
\end{align}
\end{subequations}
with clients labeled by $i=2,..,(N+1)$.

To illustrate the validity of the expansion, we consider one client ($i=2$) in a homogeneous mixture that is also composed of solvent ($i=0$) and scaffold ($i=1$).
To access when the zeroth order in the average client volume fraction ($\bar{{\phi}}_2=0$; see Sect.~\ref{app:zeroclient})is a good approximation,
we define the linearly approximated exchange activity coefficient relative to its zeroth order:
\begin{subequations}
\begin{equation}
\label{eq:delta_gamma}
    \Delta \bar{\gamma}_2 = \frac{\bar{\gamma}_2^{1}-\bar{\gamma}_2^{0}}{\bar{\gamma}_2^{0}}
    \, ,  
\end{equation}
where the first order of the client exchange activity coefficient is 
\begin{equation}
\label{eq:gamma1}
\bar{\gamma}_2^{1} = \bar{\gamma}_2^{0} +  \bar{\phi}_2 / \phi_2^{*}
\end{equation}
\end{subequations}
with $\bar{\gamma}_2^{0}={(1-{\phi}^{0}_{1})^{-r_2}}\exp{(r_2{\phi}^{0}_{1}\Delta \chi_2)}$ being the 
the zeroth order
according to Eq.~\eqref{eq:first_order_act_coeff}.
Note that $\Delta \chi_2$ is given in Eq.~\eqref{eq:abb_delta_chi}.
When decreasing the client volume fraction below the volume fraction scale,  
\begin{equation}
\label{eq:vol_scale_2}
{\phi}_2^{*}  = \left[{r_2}\frac{\exp{(r_2{\phi}^{0}_{1}\Delta \chi_2)}}{(1-{\phi}^{0}_{1})^{r_2+1}}\left(1-2(1-{\phi}^{0}_{1})\chi_{02}\right)\right]^{-1} \,
\end{equation}
taking the zeroth order of the client activity coefficient becomes a good approximation,i.e., $\bar{\gamma}_2 \simeq \bar{\gamma}_2^{0}$ for $\bar{\phi}_2 \ll \phi_2^{*}$, respectively. In this case, the activity coefficient is constant and equal to Eq.~\eqref{eq:activity_coeff_app_zero}.
Moreover,  the partition coefficient is also constant and equal to Eq.~\eqref{eq:App_partition_zero}. Fig.~\ref{fig_9}(c) confirms that below the volume scale, $\bar{\phi}_2 < \phi_2^*$, 
the relative deviation of the client activity coefficient becomes negligible, i.e., 
$\Delta \bar{\gamma}_2 = 1\%$ for $\bar{\phi}_2 = (\bar{\gamma}_2^{0}\phi_2^*)/100$.

\section{Continuum model for diluted clients}
\label{ap:cont}

Using linear response for the diffusive fluxes (Eq.~\eqref{eq:fluxes_linres}) and the mobility matrix (Eq.~\eqref{eq:mob_matrix_clients}),
the fluxes of the scaffold ($i=1$) and the clients $i=2,...,(N+1)$ are: 
\begin{align}
\textbf{j}_{1} &=-
m_{01}\phi_{1}(1-\phi_{1})
\nabla\bar{\mu}_{1}
\qc
\\
\nonumber
\textbf{j}_{i} &= -
\bigg[
m_{0i}\phi_{i}(1-\phi_{1})
+ m_{1i}\phi_{i}\phi_{1}
\bigg]
\nabla\bar{\mu}_{i}
\qd
\end{align}
We express the exchange chemical potentials for scaffold $\bar{\mu}_1$ and client $\bar{\mu}_i$ in terms of the exchange activity coefficients in the limit of vanishing client volume fraction (Eq.~\eqref{eq:activity_coeff_app_zero}), which solely depends on the scaffold volume fraction $\phi_1$:
\begin{align}
 \nabla\bar{\mu}_{1}
 &=k_{B}T\bigg[\frac{1}{\phi_{1}}+\frac{1}{\bar{\gamma_{1}}}\frac{\partial \bar{\gamma}_1}{\partial \phi_{1}}\bigg]\nabla \phi_{1} 
 - \kappa_1 \nabla \nabla^2 \phi_1
 \qc
 \\
 \nabla \bar{\mu}_{i} 
 &=k_{B}T\bigg[\frac{1}{\bar{\gamma}_{i}}\frac{\partial \bar{\gamma}_{i}}{\partial \phi_{1}}\nabla \phi_{1}+\frac{1}{\phi_{i}}\nabla \phi_{i} \bigg]
 \qc
\end{align}
where we have dropped the higher-order gradient contribution to the client flux, $-\kappa_i \nabla \nabla^2 \phi_i$.
This contribution can be neglected since the
diluted clients $i=2,..., (N+1)$ 
cannot phase-separate and do not form an interface.
Thus, the spatial transport of clients is well captured by the leading order diffusive flux that is proportional to $\nabla \phi_i$. 
Using Eq.~\eqref{eq:activity_coeff_app_zero}, we obtain Eqs.~\eqref{eq:sc_cm} and \eqref{eq:c_cm} in the main text.

\section{Condition for maximal yield for reversible chemical reactions}\label{App:4}

Here we give the solutions for reversible chemical reactions controlled by a scaffold-rich condensate and the system being at phase equilibrium. We determine the 
average product client volume fraction $\bar{\phi}_{B}(\infty)$ in the stationary state ($t\to \infty$) and compare it to the average volume fraction of the homogeneous reference system where the compartment volume is zero ($V^\I=0$).
We use such averages to obtain conditions for when the relative yield is maximal.


\subsection{Uni-molecular scheme}\label{App:4a}

For an uni-molecular reaction ($g=1$)
the average product client volume fraction reads:
\begin{widetext}
\be
\label{eq:ssuni}
\bar{\phi}_{B}(\infty)=\frac{\exp\left(\frac{\cpref_{A_{1}}}{k_BT}\right)\bar{\gamma}^\I_{A_1}P_{A_{1}}\zeta_{A_{1}}\left[\sum_{\alpha}V^{\alpha}k^{\alpha}\exp\left(\frac{\tilde{\mu}^{\alpha}_{F}}{k_BT}\right)\right]}{\exp\left(\frac{\cpref_{A_1}}{k_BT}\right)\bar{\gamma}^\I_{A_1}P_{A_{1}}\zeta_{A_{1}}\left[\sum_{\alpha}V^{\alpha}k^{\alpha}\exp\left(\frac{\tilde{\mu}_{F}^{\alpha}}{k_BT}\right)\right]+\exp\left(\frac{\cpref_{B}}{k_BT}\right)\bar{\gamma}^\I_{B}P_{B}\zeta_{B}\left[\sum_{\alpha}V^{\alpha}k^{\alpha}\right]}\psi_{1} \qc
\ee
\end{widetext}
where $\psi_{1}=(\bar{\phi}_{A_1}+\bar{\phi}_{B})$ is the conserved quantity of the uni-molecular reaction. 
For vanishing condensate volume ($V^\I=0$), the equation above simplifies:
\be
\bar{\phi}_{B}(\infty)|_{V^\I=0}=\frac{\exp\left(\frac{\cpref_{A_{1}}+\tilde{\mu}_{F}}{k_BT}\right)\bar{\gamma}_{A_1}}{\exp\left(\frac{\cpref_{A_{1}}+\tilde{\mu}_{F}}{k_BT}\right)\bar{\gamma}_{A_1}+\exp\left(\frac{\cpref_{B}}{k_BT}\right)\bar{\gamma}_{B}}\psi_{1}\qd
\ee

\begin{figure*}[tb]
\centering
\includegraphics[width=1.0\textwidth]{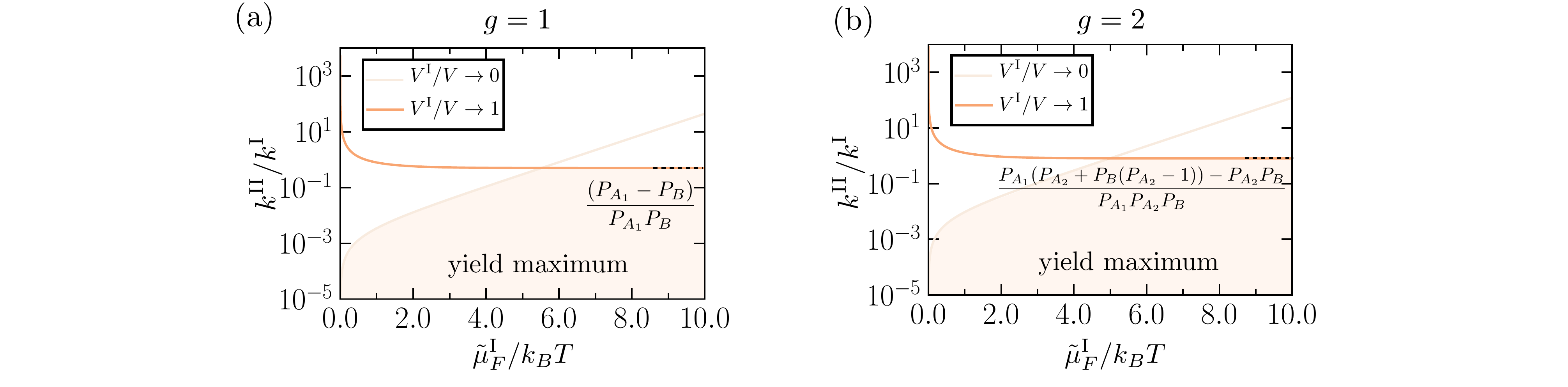}
\caption{\textbf{Graphical representation of the conditions for a maximum in product yield $\mathcal{Y}$.}
The conditions are shown for the 
uni-molecular reaction (\textbf{(a)}, $g=1$) and the bi-molecular reaction (\textbf{(b)}, $g=2$).
We vary the fraction of the reaction rate coefficients $k^\II/k^\I$ and fuel supply energy $\tilde{\mu}^\I_F$.
A yield maximum exits only in the orange shaded domain where both conditions (Eqs.~\eqref{eq:cond_uni_max} for $g=1$) and (Eqs.~\eqref{eq:reduced_conditionsg2} for $g=2$ are satisfied.}
\label{fig_si_3}
\end{figure*}

To obtain conditions for the existence of a maximum in the relative yield $\mathcal{Y}(V^\I)$ (Eq.~\eqref{eq:rel_yield_def}) as a function of the condensate volume $V^\I$, we consider the slope of the relative field, $\mathcal{Y}^{'}$.
We evaluate this slope at the limits, $V^\I\rightarrow 0$ and $V^\I\rightarrow 1$, respectively. 
If the relative yield does not have multiple extrema, the conditions of a maximum is given as:
\be
\begin{split}
&\mathcal{Y}^{'}(V^\I/V)>0 \quad\text{for}\quad  V^\I/V\rightarrow 0 \qc\\
& \mathcal{Y}^{'}(V^\I/V)<0 \quad\text{for} \quad V^\I/V\rightarrow 1 \qd
\end{split}
\ee
These conditions can be expressed as:
\be
\begin{split}\label{eq:cond_uni_max}
&\frac{k^\II}{k^\I}<\frac{\left(\exp{\left(\frac{\tilde{\mu}^{\I}_{F}}{k_BT}\right)-1}\right)}{(P_{A_1}-P_{B})} \quad\text{for}\quad  V^\I/V\rightarrow 0 \qc\\
&\frac{k^\II}{k^\I}<\frac{\exp{\left(\frac{\tilde{\mu}^{\I}_{F}}{k_BT}\right)}\left(P_{A_1}-P_{B}\right)}{P_{A_1}P_{B}\left(\exp{\left(\frac{\tilde{\mu}^{\I}_{F}}{k_BT}\right)}-1\right)} \quad\text{for}\quad  V^\I/V\rightarrow 1 \qd
\end{split}
\ee
A graphical illustration of these conditions for a varying fraction of the reaction rate coefficients $k^\II/k^\I$ and fuel supply energy $\tilde{\mu}^\I_F$ is shown in Fig.~\ref{fig_si_3}(a).  
If such conditions are satisfied, a maximum in relative yield exists at a particular value of the relative condensate volume $V^\I/V$. 
When chemical reactions are not maintained away from equilibrium ($\tilde{\mu}^{\text{I}}_{F}=0$), the conditions reduce to $k^{\text{II}}(P_{A_1}-P_{B})<0$ and $k^{\text{I}}(P_{A_1}-P_{B})>0$. Both conditions cannot be satisfied simultaneously.
This implies that for a uni-molecular reaction ($g=1$) at equilibrium, there are no maxima in the relative yield as a function of relative condensate volume $V^\I/V$ (see Fig.~\ref{fig_3uni}(c)).
However, for non-vanishing fuel energy supply $\tilde{\mu}^{\text{I}}_{F}$, the yield as a function of $V^\I/V$ can develop a pronounced maximum. This maximum already occurs at smaller volumes if more substrate $A_1$ partitions into the condensate and if the condensate phase I favors the production of the product ($k^\II/k^\I$ decreases) compared to phase II (see Fig.~\ref{fig_3uni}(c-f)).

\subsection{Bi-molecular scheme}\label{App:4b}

For the second case of a fuel-driven bi-molecular reaction ($g=2$), 
the reaction rates for the substrates $A_i$ read
\be
\begin{split}
 s^{\alpha}_{A_i}(\var) &=- \frac{s^{\alpha}_{B}(\var)}{2}\\
    &= k^{\alpha}\bigg[\exp\bigg(\frac{\bar{\mu}_{B}}{k_{B}T}\bigg)-\exp\bigg(\frac{\bar{\mu}_{A_1}+\bar{\mu}_{A_2}+\tilde{\mu}^{\alpha}_{F}}{k_{B}T}\bigg)\bigg] \qd
    \end{split}
\ee
There are two conserved quantities, i.e., $\psi_{1}=(\bar{\phi}_{A_1}+\bar{\phi}_{A_2}+\bar{\phi}_{B})$ and $\psi_{2}=(\bar{\phi}_{A_1}-\bar{\phi}_{A_2})$.
The solution at  steady state $\bar{\phi}_{B}(\infty)$ obeys the condition $\sum^{2}_{i=1}\bar{\mu}_{A_i}+\tilde{\mu}_{F}=\bar{\mu}_{B}$. Using the conserved quantities, this condition can be written as:
\begin{align}
\nonumber
\bar{\phi}_{B}(\infty)&=\exp\bigg(\frac{\cpref_{A_1}+\cpref_{A_2}-\cpref_{B}}{k_{B}T}\bigg)\bigg(\frac{\psi_{1}+\psi_{2}-\bar{\phi}_{B}(\infty)}{2}\bigg)\\
&\quad \times \bigg(\frac{\psi_{1}-\psi_{2}-\bar{\phi}_{B}(\infty)}{2}\bigg)\qd
\end{align}
The root where $0<\bar{\phi}_{B}(\infty)<1$ is
\be
\label{eq:ap_yieldbi}
\bar{\phi}_{B}(\infty) = 2C+\psi_{1}-\sqrt{4C^2+4C\psi_{1}+\psi^2_{2}} \qc
\ee
where the coefficient 
\be
\begin{split}
C&=\exp\left(\frac{\cpref_{A_1}+\cpref_{A_2}-\cpref_{B}}{k_BT}\right)\frac{\bar{\gamma}^\I_{B}P_{B}\zeta_{B}}{\prod^{2}_{i=1}\bar{\gamma}^\I_{A_{i}}P_{A_i}\zeta_{A_i}}\\
&\quad \times \left[\frac{\sum_{\alpha}V^{\alpha}k^{\alpha}}{\sum_{\alpha}V^{\alpha}k^{\alpha}\exp\left(\frac{\tilde{\mu}_{F}^{\alpha}}{k_BT}\right)}\right]
\qd
\end{split}
\ee
To solve for the existence of a maximum in the relative yield, we determine the slope of the Eq.~\eqref{eq:ap_yieldbi}.
When evaluating at,  $V^\I\rightarrow/V 0$, we find: 
\begin{widetext}
\be
\label{eq:condition_1}
\begin{split}
&\left[k^\I\left(\exp\left(\frac{\tilde{\mu}^{\I}_{F}}{k_BT}\right)-1\right)+k^\II(1-\sum_{i}P_{A_i}+P_{B})\right]\Bigg[2\exp\left(\frac{\bar{\mu}^0_{B}}{k_{B}T}\right)P_B\bar{\gamma}^{\I}_{B}+\prod^{2}_{i=1}\exp\left(\frac{\bar{\mu}^0_{A_i}}{k_{B}T}\right)P_{A_i}\bar{\gamma}^\I_{A_i}\\
&\Bigg(\psi_1-\sqrt{\frac{4\exp\left(\frac{\bar{\mu}^0_{B}-2\sum_{i}\bar{\mu}^0_{A_i}}{k_BT}\right)\left(\exp\left(\frac{\bar{\mu}^0_{B}}{k_{B}T}\right)P_B\bar{\gamma}^{\I}_{B}+\prod^{2}_{i=1}\exp\left(\frac{\bar{\mu}^0_{A_i}}{k_{B}T}\right)P_{A_i}\bar{\gamma}_{A_i}\psi_1\right)P_B\bar{\gamma}^\I_{B}}{\prod_{i}(P_{A_i}\bar{\gamma}^\I_{A_i})^2}+\psi_2^2}\Bigg)\Bigg]>0 \, , 
\end{split}
\ee
while when calculating at $V^\I/V \rightarrow 1$, we obtain:
\be
\begin{split}
\label{eq:condition_2}
& \left[k^\I\exp\left(\frac{\tilde{\mu}^{\I}_{F}}{k_BT}\right)\left((P_{A_1}(P_{A_2}+P_B(P_{A_2}-1)))-P_{A_2}P_B\right)+k^{\II} P_{A_1}P_{A_2}P_B\left(1-\exp\left(\frac{\tilde{\mu}^{\I}_{F}}{k_BT}\right)\right)\right] \\
&\Bigg[-2\exp\left(\frac{\bar{\mu}^0_{B}}{k_{B}T}\right)\bar{\gamma}^{\I}_{B}+\exp\left(\frac{\tilde{\mu}^{\I}_{F}}{k_BT}\right)\prod^{2}_{i=1}\exp\left(\frac{\bar{\mu}^0_{A_i}}{k_{B}T}\right)\bar{\gamma}_{A_i} \\
&\left(-\psi_{1}+\sqrt{\frac{4\exp\left(\frac{\bar{\mu}^0_{B}-2\sum_{i}\bar{\mu}^0_{A_i}-\tilde{\mu}^{\I}_{F}}{k_{B}T}\right)\left(\exp\left(\frac{\bar{\mu}^0_{B}-2\sum_{i}\bar{\mu}^0_{A_i}-\tilde{\mu}^{\I}_{F}}{k_{B}T}\right)\bar{\gamma}^{\I}_{B}+\prod^{2}_{i=1}\bar{\gamma}^{\I}_{A_i}\right)\bar{\gamma}^{\I}_{B}}{\prod_{i}(\bar{\gamma}^\I_{A_i})^2}+\psi^{2}_{2}}\right)\Bigg]<0 \, . 
\end{split}
\ee
For the present choice of parameters, the terms within the second round brackets in Eq.~\eqref{eq:condition_1} and Eq.~\eqref{eq:condition_2}, respectively,  is always positive. Thus, the sufficient conditions for the existence of a maximum reduce 
to: 
\begin{subequations}
\label{eq:reduced_conditionsg2}
\be
k^{\I}\left(\exp\left(\frac{\tilde{\mu}^{\I}_{F}}{\bc T}\right)-1\right)>k^{\II}\left(\sum_{i}P_{A_i}-P_{B}-1\right)\quad\text{at}\quad  V^\I/V\rightarrow 0 \, , 
\ee
\be
k^{\I}\exp\left(\frac{\tilde{\mu}^{\I}_{F}}{k_BT}\right)\left((P_{A_1}(P_{A_2}+P_B(P_{A_2}-1)))-P_{A_2}P_B\right)>k^{\II}\left(\exp\left(\frac{\tilde{\mu}^{\I}_{F}}{\bc T}\right)-1\right)
\quad\text{at}\quad  V^\I/V\rightarrow 1 \, . 
\ee
\end{subequations}
\end{widetext}
In summary, if the conditions above are satisfied, then a maximum exists as a function of condensate volume $V^\I$. 
See Fig.~\ref{fig_si_3}(b) for a graphical illustration of these conditions for varying fraction of the reaction rate coefficients, $k^\II/k^\I$, and fuel supply energy $\tilde{\mu}^\I_F$.
Interestingly, when $\tilde{\mu}^{\text{I}}_{F}=0$, the conditions above can still be satisfied for the considered case of a bi-molecular reaction ($g=2$).


\section{Analytic solutions of thin interface model for
reversible uni-molecular chemical reactions}
\label{App:thin_interface_rev_uni}

Here, we provide analytical solutions for the stationary concentration fields of diffusive clients undergoing uni-molecular reactions, with which the results of Fig.~\ref{fig_3uni}(a-d) have been produced. From the stationary case of Eq.~\eqref{eq:reacdiffbasic}, we can derive radial symmetric volume fractions for $A_1$ and $B$ in three spatial dimensions as
\begin{subequations}
\be
\begin{split}
  \phi^{\I}_{A_1}(r)&= \frac{\beta(D^\I_{A_1}C_1+D^\I_{B}C_2)}{(\alpha D^\I_{B}\exp\left(\tilde{\mu}^\I_{F}/k_{B}T\right)+\beta D^\I_{A_1})}
  \\
  &+\frac{ D^\I_{B}(\alpha C_1\exp\left(\tilde{\mu}^\I_{F}/k_{B}T\right)-\beta C_2)}{(\alpha D^\I_{B}\exp\left(\tilde{\mu}^\I_{F}/k_{B}T\right)+\beta D^\I_{A_1})}\frac{\sinh({r/l^\I})}{r/l^\I}
  \qc
\end{split}
\ee
\be
\begin{split}
  \phi^{\I}_{B}(r)&= \frac{\alpha\exp\left(\tilde{\mu}^\I_{F}/k_{B}T\right)(D^\I_{A_1}C_1+D^\I_{B}C_2)}{(\alpha D^\I_{B}\exp\left(\tilde{\mu}^\I_{F}/k_{B}T\right)+\beta D^\I_{A_1})}
  \\
  &- \frac{ D^\I_{A_1}(\alpha C_1\exp\left(\tilde{\mu}^\I_{F}/k_{B}T\right)-\beta C_2)}{(\alpha D^\I_{B}\exp\left(\tilde{\mu}^\I_{F}/k_{B}T\right)+\beta D^\I_{A_1})}\frac{\sinh({r/l^\I})}{r/l^\I} \qc
\end{split}
\ee
\be
\begin{split}
  \phi^{\II}_{A_1}(r)&= \frac{P_{B}\beta(D^\II_{A_1}(C_3+rC_4)+D^\II_{B}(C_5+rC_6))}{r(P_{A_1}\alpha D^\II_{B}+P_{B}\beta D^\II_{A_1})}
  \\
  &+
  \frac{D^\II_{B}( P_{A_1}\alpha C_3-P_{B}\beta C_5))}{(P_{A_1}\alpha D^\II_{B}+P_{B}\beta D^\II_{A_1})}\frac{\exp(-{r/l^\II})}{r}\\
  &+\frac{D^\II_{B}( P_{A_1}\alpha C_4-P_{B}\beta C_6))}{(P_{A_1}\alpha D^\II_{B}+P_{B}\beta D^\II_{A_1})}\frac{\sinh({r/l^\II})}{r/l^\II} \qc
\end{split}
\ee
\be
\begin{split}
  \phi^{\II}_{B}(r)&= \frac{P_{A_1}\alpha(D^\II_{A_1}(C_3+rC_4)+D^\II_{B}(C_5+rC_6))}{r(P_{A_1}\alpha D^\II_{B}+P_{B}\beta D^\II_{A_1})}\\
  &-
  \frac{D^\II_{B}( P_{A_1}\alpha C_3-P_{B}\beta C_5))}{(P_{A_1}\alpha D^\II_{B}+P_{B}\beta D^\II_{A_1})}\frac{\exp(-{r/l^\II})}{r}\\
  &-\frac{D^\II_{B}( P_{A_1}\alpha C_4-P_{B}\beta C_6))}{(P_{A_1}\alpha D^\II_{B}+P_{B}\beta D^\II_{A_1})}\frac{\sinh({r/l^\II})}{r/l^\II}\qc
\end{split}
\ee
\end{subequations}
where $\alpha=\exp(\bar{\mu}^{0}_{A_1}/k_{B}T)\bar{\gamma}^{\I}_{A_1}$ and $\beta=\exp(\bar{\mu}^{0}_{B}/k_{B}T)\bar{\gamma}^{\I}_{B}$. Furthermore, we have introduced the reaction-diffusion length scales
\be\label{eq:rdlengthscales}
l^{\alpha}
=\sqrt{\frac{D^{\alpha}_{A_1}D^{\alpha}_{B}}{k^{\alpha}\left[D^{\alpha}_{A_1}\exp\left(\frac{\cpref_{B}}{k_BT}\right)\bar{\gamma}^{\alpha}_{B}+D^{\alpha}_{B}\exp\left(\frac{\cpref_{A_1}+\tilde{\mu}^{\alpha}_{F}}{k_BT}\right)\bar{\gamma}^{\alpha}_{A_1}\right]}}.
\ee
Due to the linearity of Eq.~\eqref{eq:reacdiffbasic}, we can find these solutions as linear combinations of solutions of the Laplace equation and the radial symmetric spherical Bessel problem, leading in principle to eight coefficients $C_i$, which have to be determined via boundary conditions. 
However, to avoid any singularities at the droplet center, thereby fulfilling Eq.~\eqref{eq:bc_noflux_center}, we have omitted the $1/r$ solution of the Laplace equation and the Spherical Bessel function of the second kind and zeroth order ($\propto \exp(-r)/r$) in phase I. The remanding coefficients $C_1, ..., C_6$ are determined by the conditions Eq.~\eqref{eq:bc_conservation_interface}-Eq.~\eqref{eq:bc_noflux_L}.

\section{Scaling laws of monomer assembly kinetic}
\label{ap:hom_nuc}

The generic form of the dynamic equation of an irreversible assembly process with assembly order $n$ reads for the average volume fractions (definition see Eq.~\eqref{eq:avg_cli_pheq}) of the assembly $A_n$ and the monomers $A$:
\be\label{eq:nu2}
\dot{\bar{\phi}}_{A_n}(t)=k_\text{eff}\, (\psi-\bar{\phi}_{A_n}(t))^{n} = - \dot{\bar{\phi}}_{A}(t) \qd
\ee
For the case when the scaffold component phase separates with a non-zero fuel energy $\tilde{\mu}_F^\I$ in phase I, we can use the assembly rate given in Eq.~\eqref{eq:b2} and write the effective rate at phase equilibrium as:
\begin{align}
k_\text{eff}&= \frac{\nu_{A_n}}{\nu_A} (\gamma_A^\II)^n \exp(\bar{\mu}^0_A/k_B T) \zeta_{A}^n \\
 \nonumber
& \quad \times [V^I k^\I \exp(\tilde{\mu}_F^\I/k_B T) + (V-V^\I) k^\II] \qc
\end{align}
where $\zeta_{i}$ is the partitioning degree defined in Eq.~\eqref{eq:partition_degree}.
In a homogeneous system that can be achieved in our model by $V^\text{I}=0$ (implying that $\zeta_{i}=1$),
$k_\text{eff}= ({\nu_{A_n}}/{\nu_A})  (\gamma_A(\phi_1))^n \exp(\bar{\mu}^0_A/k_B T) k$, 
where $k$ denotes the assembly rate coefficient in the homogeneous system. 

The solution of the generic dynamic equation  (Eq.~\eqref{eq:nu2}) is given as: 
\be
\bar{\phi}_{A}(t)= \left((n-1)\left(k_\text{eff}\, t+\frac{\psi}{n-1}\right)\right)^{-\frac{1}{(n-1)}} \qc
\ee
where the conserved quantity of the assembly process is
$\psi=(\bar{\phi}_{A}+\bar{\phi}_{A_n})$.

\begin{figure}[tb]
\centering
\includegraphics[width=1.0\columnwidth]{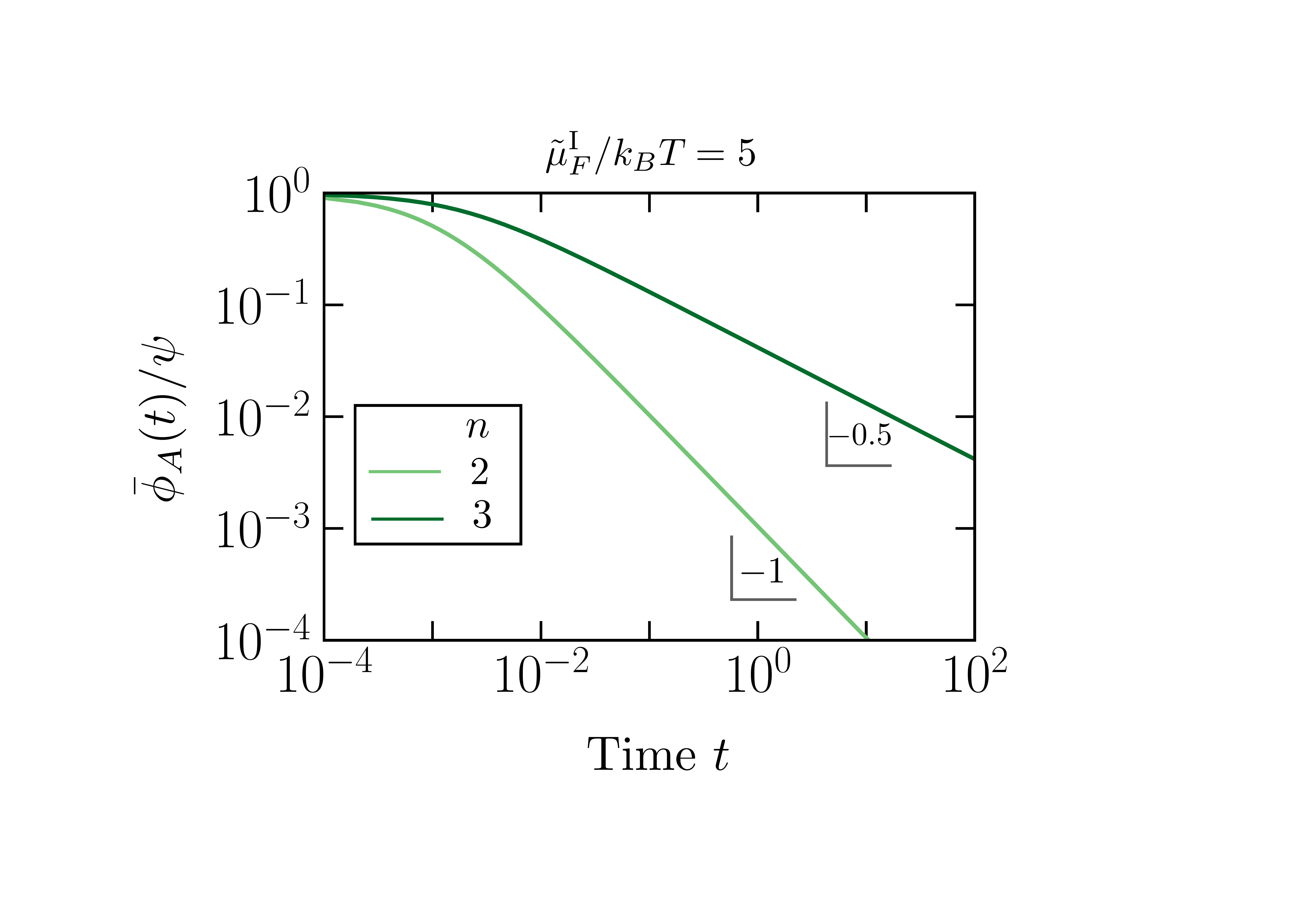}
\caption{\textbf{Long-time behavior in assembly reactions.} 
The long-time of the decreasibg monomer voluem fraction $\phi_A(t)$ is shown for the assembly orders $n=2$ and $3$. To illustrate the power-law decay, we use a log-log representation, confirming the respective power laws of $t^{-1}$ and $t^{-1/2}$. We use $\tilde{\mu}_{F}^{\I}/k_{B}T=5$ as in Fig.~\ref{fig_4}(a,b), where show a lin-lin representation of the same curve.}
\label{fig_SI_4}
\end{figure}

On long time-scales, we see that the average monomer volume fraction decays as 
$\bar{\phi}_{A}(t) \simeq t ^{-1/(n-1)}$. Specifically, for $n=2$, $\bar{\phi}_{A}(t) \simeq t ^{-1}$ while for $n=3$, 
$\bar{\phi}_{A}(t) \simeq t ^{-1/2}$.
These trends are confirmed by solving the model for the reversible assembly process in the presence of compartments (see section~\ref{sect:Reaction2}); the corresponding results are shown in 
Fig.~\ref{fig_SI_4}.
Note that for even longer times, $\bar{\phi}/\psi$ will approach a non-zero  plateau value due to the reversible pathway from the assembly $A_n$ to the monomer $A$. However, by choosing the reference chemical potential such that the assembly $A_n$ is strongly favored over the monomer $A$, i.e.,  $\exp\left(\frac{\bar{\mu}^{0}_{A_{n}}}{k_{B}T}\right) \ll \exp\left(\frac{n\bar{\mu}^{0}_{A}}{k_{B}T}\right)$,
the plateau emerges at much longer time scales as depicted in Fig.~\ref{fig_SI_4}.

\section{Parameters}

\begin{center}
\begin{tabular}{|m{0.3\columnwidth}| m{0.3\columnwidth}| } 
  \hline
  Quantities & Values \\ 
  \hline
\vtop{\hbox{\strut Fig.2b:}\hbox{\strut $r_0,r_1$} \hbox{\strut $\chi_{01}$}\hbox{\strut $\kappa_{1}$}
}  
&  \vtop{\hbox{\strut $1,1$} \hbox{\strut $3$}\hbox{\strut $15.10^{-2}$} 
}
\\ 
  \hline \vtop{\hbox{\strut  Fig.3:}\hbox{\strut $r_0,r_1,r_{A_1/A_2},r_{B}$}\hbox{\strut $\psi_1,\psi_2$ 
  } \hbox{\strut $\chi_{01},\chi_{0A_1},\chi_{0A_2},\chi_{0B}$}\hbox{\strut $P_{A_1},P_{A_2},P_{B}$}\hbox{\strut $\omega_{A_1},\omega_{A_2},\omega_{B}$}\hbox{\strut $k^\II$}\hbox{\strut $\bar{\phi}_{B}(t=0)$}
  \hbox{\strut $\tilde{D}^{\I}_{B},\tilde{D}^{\I}_{A_i},\tilde{D}^{\II}_{A_i} $}}  &  \vtop{\hbox{\strut $1,1,1,2$} \hbox{\strut $1, 0.2$} \hbox{\strut $3,2.5,2,0$}\hbox{\strut $10,10,200$}\hbox{\strut $-3.5,-3.5,1$}\hbox{\strut $0.01$}\hbox{\strut $0.01\psi_1$}
  \hbox{\strut $\frac{\tilde{D}^{\II}_{B}}{10},\frac{\tilde{D}^{\II}_{B}}{10},\frac{\tilde{D}^{\II}_{B}}{10},\tilde{D}^{\II}_{B}$}} \\
  \hline
 \vtop{\hbox{\strut Fig.4:}\hbox{\strut $r_0,r_1,r_{A_1/B}$}\hbox{\strut $\psi$  
 } \hbox{\strut $\chi_{01},\chi_{0A_1},\chi_{0B}$}\hbox{\strut $P_{A_1},P_{B}$}\hbox{\strut $\omega_{A_1},\omega_{A_2},\omega_{B}$}\hbox{\strut $k^\II$}\hbox{\strut $\bar{\phi}_{B}(t=0)$}
 \hbox{\strut $\tilde{D}^{\I}_{B},\tilde{D}^{\I}_{A_1},\tilde{D}^{\II}_{A_1} $}}  &  \vtop{\hbox{\strut $1,1,1$} \hbox{\strut $1$} \hbox{\strut $3,2.5,2$}\hbox{\strut $500,2$}\hbox{\strut $-4,0$}\hbox{\strut $0.01$}\hbox{\strut $10^{-4}\psi$}
 \hbox{\strut $\frac{\tilde{D}^{\II}_{B}}{10},\frac{3}{10}\tilde{D}^{\II}_{B},3\tilde{D}^{\II}_{B}$}} \\
  \hline
  \vtop{\hbox{\strut \quad \,  Fig.5:}\hbox{\strut \quad \,  $r_0,r_1,r_{A},r_{A_2},r_{A_3}$}\hbox{\strut \quad \,   $\psi$ 
  } \hbox{\strut \quad \,   $\chi_{01},\chi_{0A},\chi_{0A_2},\chi_{0A_3}$}\hbox{\strut \quad \,   $P_{A},P_{A_2},P_{A_3}$}\hbox{\strut \quad \,   $\omega_{A},\omega_{A_2},\omega_{A_3}$}\hbox{\strut \quad \,   $k^\II$}\hbox{\strut \quad \,   $\bar{\phi}_{A_{2/3}}(t=0)$}
  \hbox{\strut \quad \,   $\tilde{D}^{\I}_{A_2},\tilde{D}^{\I}_{A},\tilde{D}^{\II}_{A} $}}  &  \vtop{\hbox{\strut $1,1,1,2,3$} \hbox{\strut $10^{-4}$} \hbox{\strut $3,2.5,2,2$}\hbox{\strut $10,10,10$}\hbox{\strut $5.5,-7,-5$}\hbox{\strut $0.01$}\hbox{\strut $10^{-2}\psi$}
  \hbox{\strut $\frac{\tilde{D}^{\II}_{A_2}}{10},\frac{\tilde{D}^{\II}_{A_2}}{5},2\tilde{D}^{\II}_{A_2}$}} \\
  \hline
  \vtop{\hbox{\strut Fig.6:} \hbox{\strut $r_0,r_1,r_{2}$}\hbox{\strut $\chi_{01},\chi_{02},\chi_{12}$} 
  }  &  \vtop{\hbox{\strut $1,1,1$}\hbox{\strut $3,-1,0$}} \\
  \hline
  \vtop{\hbox{\strut Fig.7:} \hbox{\strut Parameters}}  &  \vtop{\hbox{\strut Same as Figs.~\ref{fig_3uni} and \ref{fig_3}, respectively}} \\
  \hline
   \vtop{\hbox{\strut Fig.8:} \hbox{\strut Parameters}}  &  \vtop{\hbox{\strut Same as Figs.~\ref{fig_4}(a) and (b)}} \\
  \hline
\end{tabular}
\end{center}

\newpage
\cleardoublepage


\begin{thebibliography}{10}

\bibitem{Wilson:1899}
Edmund~B. Wilson.
\newblock The structure of protoplasm.
\newblock {\em Science}, 10(237):33--45, 1899.

\bibitem{Hyman:2014}
Anthony~A. Hyman, Christoph~A. Weber, and Frank J\"{u}licher.
\newblock Liquid-liquid phase separation in biology.
\newblock {\em Annual Review of Cell and Developmental Biology}, 30(1):39--58,
  2014.
\newblock PMID: 25288112.

\bibitem{Shin:2017}
Yongdae Shin and Clifford~P. Brangwynne.
\newblock Liquid phase condensation in cell physiology and disease.
\newblock {\em Science}, 357, 2017.

\bibitem{Brangwynne:2009}
Clifford~P. Brangwynne, Christian~R. Eckmann, David~S. Courson, Agata Rybarska,
  Carsten Hoege, J\"{o}bin Gharakhani, Frank J{\"u}licher, and Anthony~A. Hyman.
\newblock Germline p granules are liquid droplets that localize by controlled
  dissolution/condensation.
\newblock {\em Science}, 324(5935):1729--1732, 2009.

\bibitem{Feric:2016}
Marina Feric, Nilesh Vaidya, Tyler~S. Harmon, Diana~M. Mitrea, Lian Zhu,
  Tiffany~M. Richardson, Richard~W. Kriwacki, Rohit~V. Pappu, and Clifford~P.
  Brangwynne.
\newblock Coexisting liquid phases underlie nucleolar subcompartments.
\newblock {\em Cell}, 165(7):1686--1697, 2016.

\bibitem{Boeynaems:2018}
Steven Boeynaems, Simon Alberti, Nicolas~L. Fawzi, Tanja Mittag, Magdalini
  Polymenidou, Frederic Rousseau, Joost Schymkowitz, James Shorter, Benjamin
  Wolozin, Ludo {Van Den Bosch}, Peter Tompa, and Monika Fuxreiter.
\newblock Protein phase separation: A new phase in cell biology.
\newblock {\em Trends in Cell Biology}, 28(6):420--435, 2018.

\bibitem{Antifeeva:2022}
Iuliia~A. Antifeeva, Alexander~V. Fonin, Anna~S. Fefilova, Olesya~V.
  Stepanenko, Olga~I. Povarova, Sergey~A. Silonov, Irina~M. Kuznetsova,
  Vladimir~N. Uversky, and Konstantin~K. Turoverov.
\newblock Liquid--liquid phase separation as an organizing principle of
  intracellular space: overview of the evolution of the cell
  compartmentalization concept.
\newblock {\em Cellular and Molecular Life Sciences}, 79(5):251, Apr 2022.

\bibitem{Banani:2017}
Salman~F Banani, Hyun~O Lee, Anthony~A Hyman, and Michael~K Rosen.
\newblock Biomolecular condensates: organizers of cellular biochemistry.
\newblock {\em Nature Reviews Molecular Cell Biology}, 18(5):285--298, May
  2017.

\bibitem{Alberti:2017}
Simon Alberti.
\newblock Phase separation in biology.
\newblock {\em Current Biology}, 27(20):R1097--R1102, 2017.

\bibitem{PEDERSEN:1987}
Peter~L. Pedersen and Ernesto Carafoli.
\newblock Ion motive atpases. i. ubiquity, properties, and significance to cell
  function.
\newblock {\em Trends in Biochemical Sciences}, 12:146--150, 1987.

\bibitem{Beyenbach:2006}
Klaus~W. Beyenbach and Helmut Wieczorek.
\newblock The v-type h+ atpase: molecular structure and function,physiological
  roles and regulation.
\newblock {\em Journal of Experimental Biology}, 209(4):577--589, Feb 2006.

\bibitem{HEALD:2014}
Rebecca Heald and Orna Cohen-Fix.
\newblock Morphology and function of membrane-bound organelles.
\newblock {\em Current Opinion in Cell Biology}, 26:79--86, 2014.
\newblock Cell architecture.

\bibitem{DITLEV:2018}
Jonathon~A. Ditlev, Lindsay~B. Case, and Michael~K. Rosen.
\newblock Who's in and who's out?compositional control of biomolecular
  condensates.
\newblock {\em Journal of Molecular Biology}, 430(23):4666--4684, 2018.
\newblock Phase Separation in Biology and Disease.

\bibitem{Frankel:2016}
Erica~A. Frankel, Philip~C. Bevilacqua, and Christine~D. Keating.
\newblock Polyamine/nucleotide coacervates provide strong compartmentalization
  of mg2+, nucleotides, and rna.
\newblock {\em Langmuir}, 32(8):2041--2049, 2016.
\newblock PMID: 26844692.

\bibitem{WOODRUFF:2017}
Jeffrey~B. Woodruff, Beatriz {Ferreira Gomes}, Per~O. Widlund, Julia Mahamid,
  Alf Honigmann, and Anthony~A. Hyman.
\newblock The centrosome is a selective condensate that nucleates microtubules
  by concentrating tubulin.
\newblock {\em Cell}, 169(6):1066--1077.e10, 2017.

\bibitem{AUMILLER:2017}
William~M. Aumiller and Christine~D. Keating.
\newblock Experimental models for dynamic compartmentalization of biomolecules
  in liquid organelles: Reversible formation and partitioning in aqueous
  biphasic systems.
\newblock {\em Advances in Colloid and Interface Science}, 239:75--87, 2017.
\newblock Complex Coacervation: Principles and Applications.

\bibitem{Elbaum-Garfinkle:2015}
Shana Elbaum-Garfinkle, Younghoon Kim, Krzysztof Szczepaniak, Carlos
  Chih-Hsiung Chen, Christian~R Eckmann, Sua Myong, and Clifford~P Brangwynne.
\newblock The disordered {P} granule protein {LAF-1} drives phase separation
  into droplets with tunable viscosity and dynamics.
\newblock {\em Proc Natl Acad Sci U S A}, 112(23):7189--7194, May 2015.

\bibitem{Lars:2021}
Lars Hubatsch, Louise~M Jawerth, Celina Love, Jonathan Bauermann, TY~Dora Tang,
  Stefano Bo, Anthony~A Hyman, and Christoph~A Weber.
\newblock Quantitative theory for the diffusive dynamics of liquid condensates.
\newblock {\em eLife}, 10:e68620, oct 2021.

\bibitem{Strulson:2012}
Christopher~A. Strulson, Rosalynn~C. Molden, Christine~D. Keating, and
  Philip~C. Bevilacqua.
\newblock Rna catalysis through compartmentalization.
\newblock {\em Nature Chemistry}, 4(11):941--946, Nov 2012.

\bibitem{Sokolova:2013}
Ekaterina Sokolova, Evan Spruijt, Maike M.~K. Hansen, Emilien Dubuc, Joost
  Groen, Venkatachalam Chokkalingam, Aigars Piruska, Hans~A. Heus, and Wilhelm
  T.~S. Huck.
\newblock Enhanced transcription rates in membrane-free protocells formed by
  coacervation of cell lysate.
\newblock {\em Proceedings of the National Academy of Sciences},
  110(29):11692--11697, 2013.

\bibitem{Drobot:2018}
Bj{\"o}rn Drobot, Juan~M. Iglesias-Artola, Kristian Le~Vay, Viktoria Mayr,
  Mrityunjoy Kar, Moritz Kreysing, Hannes Mutschler, and T-Y~Dora Tang.
\newblock Compartmentalised rna catalysis in membrane-free coacervate
  protocells.
\newblock {\em Nature Communications}, 9(1):3643, Sep 2018.

\bibitem{nakashima:2018}
Karina~K. Nakashima, Jochem~F. Baaij, and Evan Spruijt.
\newblock Reversible generation of coacervate droplets in an enzymatic network.
\newblock {\em Soft Matter}, 14:361--367, 2018.

\bibitem{arosio:2021}
Andreas~M. Kueffner, Miriam Linsenmeier, Fulvio Grigolato, Marc Prodan, Remo
  Zuccarini, Umberto Capasso~Palmiero, Lenka Faltova, and Paolo Arosio.
\newblock Sequestration within biomolecular condensates inhibits abeta 42
  amyloid formation.
\newblock {\em Chem. Sci.}, 12:4373--4382, 2021.

\bibitem{O?Flynn:2021}
Brian~G. O'Flynn and Tanja Mittag.
\newblock A new phase for enzyme kinetics.
\newblock {\em Nature Chemical Biology}, 17(6):628--630, Jun 2021.

\bibitem{Schoenmakers:2023}
Ludo L.~J. Schoenmakers, N.~Amy Yewdall, Tiemei Lu, Alain A.~M. Andr{\'e},
  Frank.~H.T. Nelissen, Evan Spruijt, and Wilhelm T.~S. Huck.
\newblock In vitro transcription--translation in an artificial biomolecular
  condensate.
\newblock {\em ACS Synthetic Biology}, Jun 2023.

\bibitem{weber2019physics}
Christoph~A Weber, David Zwicker, Frank J{\"u}licher, and Chiu~Fan Lee.
\newblock Physics of active emulsions.
\newblock {\em Reports on Progress in Physics}, 82(6):064601, 2019.

\bibitem{Bo:2021}
Stefano Bo, Lars Hubatsch, Jonathan Bauermann, Christoph~A. Weber, and Frank
  J\"ulicher.
\newblock Stochastic dynamics of single molecules across phase boundaries.
\newblock {\em Phys. Rev. Research}, 3:043150, Dec 2021.

\bibitem{zwicker2022intertwined}
David Zwicker.
\newblock The intertwined physics of active chemical reactions and phase
  separation.
\newblock {\em Current Opinion in Colloid \& Interface Science}, page 101606,
  2022.

\bibitem{BauermannLaha:2022}
Jonathan Bauermann, Sudarshana Laha, Patrick~M. McCall, Frank J{\"u}licher, and
  Christoph~A. Weber.
\newblock Chemical kinetics and mass action in coexisting phases.
\newblock {\em Journal of the American Chemical Society}, 144(42):19294--19304,
  Oct 2022.

\bibitem{Bauermann:2022}
Jonathan Bauermann, Christoph~A. Weber, and Frank J{\"u}licher.
\newblock Energy and matter supply for active droplets.
\newblock {\em Annalen der Physik}, 534(9):2200132, 2022.

\bibitem{ditlev2018s}
Jonathon~A Ditlev, Lindsay~B Case, and Michael~K Rosen.
\newblock Who's in and who's out?compositional control of biomolecular
  condensates.
\newblock {\em Journal of molecular biology}, 430(23):4666--4684, 2018.

\bibitem{gao2022brief}
Yifei Gao, Xi~Li, Pilong Li, and Yi~Lin.
\newblock A brief guideline for studies of phase-separated biomolecular
  condensates.
\newblock {\em Nature Chemical Biology}, 18(12):1307--1318, 2022.

\bibitem{woodruff2018organization}
Jeffrey~B Woodruff, Anthony~A Hyman, and Elvan Boke.
\newblock Organization and function of non-dynamic biomolecular condensates.
\newblock {\em Trends in biochemical sciences}, 43(2):81--94, 2018.

\bibitem{Lyon:2021}
Andrew~S. Lyon, William~B. Peeples, and Michael~K. Rosen.
\newblock A framework for understanding the functions of biomolecular
  condensates across scales.
\newblock {\em Nature Reviews Molecular Cell Biology}, 22(3):215--235, Mar
  2021.

\bibitem{Frank_JProst2008}
Frank J{\"u}licher and Jacques Prost.
\newblock Generic theory of colloidal transport.
\newblock {\em The European physical journal. E, Soft matter}, 29:27--36, 05
  2009.

\bibitem{Bray:1993}
Alan~J. Bray.
\newblock Theory of phase-ordering kinetics.
\newblock {\em Advances in Physics}, 51:481 -- 587, 1993.

\bibitem{elder2001sharp}
KR~Elder, Martin Grant, Nikolas Provatas, and JM~Kosterlitz.
\newblock Sharp interface limits of phase-field models.
\newblock {\em Physical Review E}, 64(2):021604, 2001.

\bibitem{weber2019spatial}
Christoph Weber, Thomas Michaels, and L~Mahadevan.
\newblock Spatial control of irreversible protein aggregation.
\newblock {\em Elife}, 8:e42315, 2019.

\bibitem{michaels2022enhanced}
Thomas~CT Michaels, L~Mahadevan, and Christoph~A Weber.
\newblock Enhanced potency of aggregation inhibitors mediated by liquid
  condensates.
\newblock {\em Physical Review Research}, 4(4):043173, 2022.

\bibitem{milo:2015}
Ron Milo and Rob Philips.
\newblock {\em Cell biology by the numbers}.
\newblock CRC Press, 2015.

\bibitem{du2023condensate}
Zhenzhen Du, Kun Shi, Jordan~S Brown, Tao He, Wei-Sheng Wu, Ying Zhang,
  Heng-Chi Lee, and Donglei Zhang.
\newblock Condensate cooperativity underlies transgenerational gene silencing.
\newblock {\em Cell reports}, 42(8), 2023.

\bibitem{hernandez2017local}
Amayra Hern{\'a}ndez-Vega, Marcus Braun, Lara Scharrel, Marcus Jahnel, Susanne
  Wegmann, Bradley~T Hyman, Simon Alberti, Stefan Diez, and Anthony~A Hyman.
\newblock Local nucleation of microtubule bundles through tubulin concentration
  into a condensed tau phase.
\newblock {\em Cell reports}, 20(10):2304--2312, 2017.

\bibitem{patel2015liquid}
Avinash Patel, Hyun~O Lee, Louise Jawerth, Shovamayee Maharana, Marcus Jahnel,
  Marco~Y Hein, Stoyno Stoynov, Julia Mahamid, Shambaditya Saha, Titus~M
  Franzmann, et~al.
\newblock A liquid-to-solid phase transition of the als protein fus accelerated
  by disease mutation.
\newblock {\em Cell}, 162(5):1066--1077, 2015.

\bibitem{Fares:2020}
Hadi~M. Fares, Alexander~E. Marras, Jeffrey~M. Ting, Matthew~V. Tirrell, and
  Christine~D. Keating.
\newblock Impact of wet-dry cycling on the phase behavior and
  compartmentalization properties of complex coacervates.
\newblock {\em Nature Communications}, 11(1):5423, Oct 2020.

\bibitem{Tekin:2022}
Emre Tekin, Annalena Salditt, Philipp Schwintek, Sreekar Wunnava, Juliette
  Langlais, James Saenz, Dora Tang, Petra Schwille, Christof Mast, and Dieter
  Braun.
\newblock Prebiotic foam environments to oligomerize and accumulate rna.
\newblock {\em ChemBioChem}, 23(24):e202200423, 2022.

\bibitem{haugerud2024nonequilibrium}
Ivar~Svalheim Haugerud, Pranay Jaiswal, and Christoph~A Weber.
\newblock Nonequilibrium wet--dry cycling acts as a catalyst for chemical
  reactions.
\newblock {\em The Journal of Physical Chemistry B}, 2024.

\bibitem{mutschler2015freeze}
Hannes Mutschler, Aniela Wochner, and Philipp Holliger.
\newblock Freeze--thaw cycles as drivers of complex ribozyme assembly.
\newblock {\em Nature Chemistry}, 7(6):502--508, 2015.

\bibitem{Barto:2023}
Giacomo Bartolucci, Adriana~Calaca Serrao, Philipp Schwintek, Alexandra
  K\"{u}hnlein, Yash Rana, Philipp Janto, Dorothea Hofer, Christof~B. Mast, Dieter
  Braun, and Christoph~A. Weber.
\newblock Sequence self-selection by cyclic phase separation.
\newblock {\em Proceedings of the National Academy of Sciences},
  120(43):e2218876120, 2023.

\bibitem{bartolucci2023interplay}
Giacomo Bartolucci, Ivar~S. Haugerud, Thomas~CT Michaels, and Christoph~A
  Weber.
\newblock The interplay between biomolecular assembly and phase separation.
\newblock {\em bioRxiv}, pages 2023--04, 2023.

\bibitem{epstein2016reaction}
Irving~R Epstein and Bing Xu.
\newblock Reaction--diffusion processes at the nano-and microscales.
\newblock {\em Nature nanotechnology}, 11(4):312--319, 2016.

\bibitem{halatek2018rethinking}
Jacob Halatek and Erwin Frey.
\newblock Rethinking pattern formation in reaction--diffusion systems.
\newblock {\em Nature Physics}, 14(5):507--514, 2018.

\bibitem{menou2023physical}
Lucas Menou, Chengjie Luo, and David Zwicker.
\newblock Physical interactions in non-ideal fluids promote turing patterns.
\newblock {\em Journal of the Royal Society Interface}, 20(204):20230244, 2023.

\bibitem{ponisch2023aggregation}
Wolfram P{\"o}nisch, Thomas~CT Michaels, and Christoph~A Weber.
\newblock Aggregation controlled by condensate rheology.
\newblock {\em Biophysical Journal}, 122(1):197--214, 2023.

\bibitem{Lahathe:2023}
Sudarshana Laha.
\newblock {\em Chemical reactions controlled through compartmentalization:
  Applications to bottom-up design of synthetic life}.
\newblock monograph, Technische Universit\"{a}t Dresden, June 2023.

\end{thebibliography}

\end{document}